# Partitioning of Eddy Covariance Footprint Evapotranspiration Using Field Data, UAS Observations and GeoAI in the U.S. Chihuahuan Desert

Habibur R. Howlider[a], Hernan A. Moreno[a,*], Marguerite E. Mauritz[b], Stephanie N. Marquez[a]

[a]*Department of Earth, Environmental and Resource Sciences, University of Texas at El Paso, 500 W. University Ave, El Paso, TX 79968, USA*

[b]*Department of Biological Sciences, University of Texas at El Paso, 500 W. University Ave, El Paso, TX 79968, USA*

**Abstract**

This study proposes a new method for computing transpiration across an eddy covariance footprint using field observations of plant sap flow, phytomorphology sampling, uncrewed aerial system (UAS), deep learning-based digital image processing, and eddy covariance micrometeorological measurements. The method is applied to the Jornada Experimental Range, New Mexico, where we address three key questions: (1) What are the daily summer transpiration rates of Mesquite (*Prosopis glandulosa*) and Creosote (*Larrea tridentata*) individuals, and how do these species contribute to footprint-scale evapotranspiration? (2) How can the plant-level measurements be integrated for terrain-wide transpiration estimates? (3) What is the contribution of transpiration to total evapotranspiration within the eddy covariance footprint? Data collected from June to October 2022, during the North American Monsoon season, include hourly evapotranspiration and precipitation rates from the Ameriflux eddy covariance system (US Jo-1 Bajada site) and sap flux rates from heat-balance sensors. We used plant biometric measurements and supervised classification of multispectral imagery to upscale from the patch to footprint-scale estimations. A proportional relationship between the plant's horizontal projected area and the estimated number of water flow conduits was extended to the eddy covariance footprint via UAS data. Our results show that Mesquite's average daily summer transpiration is 2.84 mm/d, while Creosote's is 1.78 mm/d (a ratio of 1.6:1). The summer footprint integrated transpiration to evapotranspiration ratio (T/ET) was 0.50, decreasing to 0.44 during dry spells and increasing to 0.63 following significant precipitation. Further testing of this method is needed in different regions to validate its applicability. With appropriate adjustments, it could be relevant for other areas with similar ecological conditions.

[*]Corresponding author. Geological Sciences Building Room 321-A, El Paso, TX, 79968, USA.
Email: moreno@utep.edu (H.Moreno), hrhowlider@miners.utep.edu (H.Howlider), memauritz@utep.edu (M.Mauritz), snmarquez@miners.utep.edu (S.Marquez)

**1. Introduction**

Partitioning evapotranspiration (ET) is crucial for comprehensively understanding the water balance across diverse ecosystems, including those in arid conditions, complex terrain, or under intense anthropogenic influence (Scott et al., 2021). By discerning the individual contributions of evaporation (E) from soil and wetted surfaces and transpiration (T) from vegetation, researchers can better parameterize land surface-atmosphere models and understand the relationships between soil moisture, atmospheric demand, carbon fluxes, and stocks.

ET partitioning is particularly interesting in arid and semi-arid regions where water scarcity is a pressing issue and vegetation succession is ongoing (Scott and Biederman, 2017). While ET in dryland ecosystems has been successfully estimated using the Bowen Ratio (Dugas et al., 1998; Kurc and Small, 2004; Scott et al., 2006b) and open path Eddy Covariance techniques, the effects of changes in ecohydrological processes overtime on the partitioning of ET fluxes (e.g. Creosote bush expansion), remain poorly understood (Loik et al., 2004; Potts et al., 2006). These ecohydrological processes include shifts in vegetation composition and structure, variability in rainfall patterns and infiltration depths, the presence of restrictive soil layers such as caliche that limit deep percolation, and the dynamic response of plant transpiration to pulse precipitation events. Each of these factors influences how water is stored, accessed, and lost from the soil-plant-atmosphere system, ultimately affecting the balance between evaporation and transpiration components of ET.

Previous studies have suggested that combining micro-meteorological (Bowen ratio or eddy covariance data), eco-physiological (sap-flow or isotopic measurements), hydrological (micro-lysimeters, tensiometers), and high-resolution remote sensing methods would allow for ET partitioning approximations (Williams et al., 2004; Yepez et al., 2005; Scott et al., 2006a). While numerous techniques have been developed for partitioning evapotranspiration (ET), generalizing individual plant-level transpiration (T) measurements across larger spatial domains, such as eddy covariance (EC) footprints or ecosystem regions, remains a major challenge. This difficulty is especially pronounced in heterogeneous dryland ecosystems where vegetation composition, structure, and water use strategies vary considerably. The absence of widely adopted, generalizing methods limits our ability to represent community-scale water fluxes based on point-level data. As discussed in previous studies (e.g., Williams et al. (2004); Scott and Biederman (2017), integrating micrometeorological, ecophysiological, and remote sensing methods holds promise for addressing this gap. In this study, we present an approach that combines species-specific sap flow measurements, detailed plant biometric data, and high-resolution UAS-based vegetation classification to scale transpiration estimates to the EC footprint. This framework aims to reduce scaling uncertainties by accounting for both inter- and intra-species variability and actual spatial vegetation distributions.

Since most summer precipitation (P) events in dryland ecosystems are relatively few and modest in the northern Chihuahuan desert (i.e. southwestern USA; Petrie et al. (2014)), only the top few centimeters of the soil are typically saturated after rain showers, and the water is quickly consumed by soil evaporation (E) and shallow-root plant transpiration (T) due to the high atmospheric demand for water (Carlson et al., 1990; Huxman et al., 2005; Scott, 2010). However, deeper soil moisture infiltrated after stratiform winter and spring precipitation or intense, but infrequent, summer convective thundershowers is associated with higher contributions of T (Kurc and Small, 2007). Previous research has shown that T/ET tends to be high after isolated, individual intense summer precipitation events (Schlesinger and Jasechko, 2014; Sun et al., 2019). Another factor that controls the partitioning of ET is the depth of the soil horizon. The presence of indurated and spatially-continuous caliche ($CaCO_3$) layers limits deep (>1 m) water flow (Duniway et al., 2009; Nobles et al., 2010), which favors shallow water accumulation (above and around the root-zone layer) and, therefore, vegetation water use for long periods (Szutu and Papuga, 2019). In synthesis, despite other minor contributing factors like terrain slope and soil type, in arid and semi-arid regions, the frequency and strength of the precipitation pulses, the presence of a calcium carbonate horizon, and the spatial distribution and type of plant individuals affect the vertical distribution of soil moisture and consequently, the partitioning of ET (Marion and Fonteyn, 1990; Kraimer et al., 2005).

Inputs and outputs of water (e.g., precipitation P, evaporation E, transpiration T, and runoff R) are intrinsically tied to the soil–vegetation carbon budget in dryland ecosystems. These fluxes share temporal dynamics: abiotic water loss through E occurs rapidly after precipitation events, while biotic responses such as T and plant carbon uptake lag behind. This temporal decoupling mirrors what occurs in the carbon cycle. In the minutes following rainfall, there is a piston-like burst of volatile carbon release from the soil (Deng et al., 2017; Gabriel and Kellman, 2014), followed by a slower, moisture-dependent increase in vegetation carbon assimilation via net primary productivity (Huxman et al., 2005). Thus, the partitioning of ET is not only critical for quantifying water use but also for linking short- and long-term carbon flux responses in arid and semi-arid systems. Unfortunately, data on the partitioning of ET and carbon exchange characteristics at an adequate temporal and regional scale are insufficient to understand these critical ecosystem transient states (Scott et al., 2006a), and this is why information on how and when plants use soil moisture is still required (Cavanaugh et al., 2011; Potts et al., 2006).

While a number of empirical studies have measured T/ET in semiarid shrublands, including those compiled by Schlesinger and Jasechko (2014), such measurements often remain limited in spatial or temporal coverage due to the high cost and logistical complexity associated with instrumentation and data acquisition (Moran et al., 2009). Long-term T/ET ratios in drylands are typically lower than 0.5, but in wetter climates, they can reach up to 0.7 (Bu et al., 2024). In temperate and boreal ecosystems, studies using isotopic and sap flow approaches have found that T/ET can exceed 0.7 during peak growing seasons when canopy cover and soil moisture are high (Jasechko et al., 2013; Sutanto et al., 2014) indicating strong transpiration dominance under favorable conditions. However, daily rates by species individuals have rarely been estimated. Ferretti et al. (2003) conducted a study in a semiarid northeastern Colorado shortgrass steppe from May 1999 to October 2001 using a basic isotopic mass balance technique. The sum of total T and E losses was comparable to the actual ET determined from a nearby Bowen ratio energy balance system. The amount of water lost by evaporation (E/ET) varied depending on when precipitation inputs were received; it ranged from zero to roughly 40% during the growing season and up to 90% during the dormant season. Lascano et al. (1992) used commercially available gauges to determine the daily and seasonal water use of three-year-old Chardonnay plants in New Deal, Texas, to evaluate the applicability of the stem heat balancing method. According to their findings, the stem flow gauges' accuracy ranged from 5% to 10% of the daily transpiration amount as determined by gravimetric measurements. Inter-plant transpiration variability was significantly decreased when the area of its leaves was used to normalize water flux and find the total sap flow of each plant. Scott et al. (2006b) measured whole plant transpiration using the heat balance sap flow method, while evapotranspiration and net ecosystem exchange (NEE) of $CO_2$ were quantified using the Bowen ratio technique. They discovered that E exceeded T at the onset of the rainy season. E peaked and began to fall quickly after rain events once the rain started, while T often reached its peak hours (or sometimes days) following E and began to decrease subsequently, with T values proportional to the size of the precipitation pulse. An overarching observation is the lack of generalizing methods for extending individual plant T measurements over larger areas, like those of eddy covariance footprints or ecosystem regions, where the individual observations can be used as a representative of the entire community dynamics. While this discussion centers on transpiration, it is important to acknowledge that evaporation, including soil evaporation and potential contributions from canopy interception, also plays a significant role in the overall ET balance. Leaf interception was not explicitly considered here and may be underrepresented, particularly given that data from rainy periods are typically excluded in open-path eddy covariance systems due to sensor limitations during precipitation events.

Although no broadly applicable or standardized method currently exists, several studies have developed case-specific approaches to scale individual plant transpiration. For instance, Dukat et al. (2023) quantified stand-level transpiration in Scots pine forests in Poland using sap flow and biometric scaling. Similarly, Javadian et al. (2024) combined high-resolution thermal imaging with individual sap flow data to explore species-specific drought responses in mixed forests. In arid systems, Oishi et al. (2008) scaled sap flux measurements in southeastern U.S. forests using sapwood area relationships, while Cavanaugh et al. (2011) applied isotopic and sap flow data to estimate T/ET in dryland mesquite ecosystems. Alam et al. (2021) also demonstrated a canopy-level transpiration estimation by scaling stomatal conductance with LAI and intercepted radiation. These examples illustrate that while site- and species-specific extrapolations are feasible, a universal methodology remains a gap in ecohydrological science.

The objectives of this study are to (1) find out how the dominant Mesquite *(Prosopis glandulosa)* and Creosote bush *(Larrea tridentata)* vegetation species, typical of the dryland southwestern United States, use water under different summer weather conditions and to (2) propose and test a sapflux and remote sensing based method to scale up T/ET measurements over an eddy covariance footprint. This research uses a sap flux network of sensors installed in small and large Mesquite and Creosote individuals, representative of the landscape. Branch counts at a consistent vegetation height were used to establish a biometric relationship between canopy surface area and branch number, which serves as the basis to scale individual plant transpiration (T) to footprint-scale activity using UAS multi-spectral imagery. Auxiliary variables (i.e., ET and P) from a nearby micro-meteorological station are also used. Differences in transpiration rates between Creosote and Mesquite are likely influenced by plant-specific traits such as root depth, branch architecture, regional rainfall patterns, and water use efficiency, defined as carbon gained per unit of water lost. These factors have been shown to mediate how desert shrubs respond to hydrological variability, contributing to their niche differentiation in arid environments (Donovan and Ehleringer, 1991; Schenk and Jackson, 2002; Smith and Allen, 1996). A deeper understanding of these mechanisms enhances our capacity to predict vegetation responses under changing precipitation regimes (Huxman et al., 2005; Reynolds et al., 2004). With a deeper understanding of plant physiological responses to precipitation variability and soil moisture dynamics, such as changes in root uptake strategies, stomatal regulation, and water use efficiency, it becomes possible to more accurately model ecosystem carbon and water fluxes. This mechanistic insight enables clearer attribution of observed changes in vegetation composition and ecosystem function to global climate drivers, thereby improving our predictions of vegetation succession under environmental change. Moreover, the findings contribute to the refinement of land surface and hydrological models by offering empirical data for calibrating evapotranspiration (ET) partitioning. While this research is grounded in semiarid ecosystems, the methods and insights have broader relevance across water-limited biomes and can be adapted for comparative use in tropical drylands, Mediterranean systems, and temperate grasslands. Ultimately, these contributions are expected to enhance the generalizability and reliability of eddy covariance-derived flux measurements at multiple spatial scales.

## 2. Methodology

### 2.1. Study Site

This research was conducted at the Jornada Experimental Range (JER) within the eddy covariance tower footprint of the Ameriflux Bajada (US-Jo1) site in southern New Mexico (USA) in the northern range of the Chihuahuan Desert (Figure 1). Roughly, the JER comprises 200,000 hectares of land within New Mexico's Dona Ana County. Within the study site (Figure 1a), vegetation is dominated by mixed shrublands, including honey Mesquite *(Prosopis Glandulosa)* and Creosote shrub *(Larrea Tridentata)* (Serna Pérez et al., 2006; Bergametti and Gillette, 2010). Outside the study area, desert grasslands intermix with the shrubs (McClaran and Van Devender, 1997). The JER climate is characterized by a mean annual air temperature of 15 °C and an average annual precipitation of 233 mm (1991–2020) (Browning et al., 2017). The lowest annual precipitation on record was 77.0 mm in 1953, and the highest was 507.2 mm in 1984. During the North American Monsoon season (July–September), temperature and precipitation patterns vary markedly across months. July marks the onset with high daytime temperatures (33–40°C) and increasing but scattered rainfall. August is typically the peak, bringing the highest rainfall, often 30–40% of annual totals, along with moderated temperatures (30–36°C) due to cloud cover. September sees a decline in rainfall and a gradual cooling (28–33°C), signaling the retreat of monsoonal activity. This intra-seasonal variability influences plant water use and ecosystem fluxes, making it crucial for interpreting seasonal patterns in land-atmosphere interactions. Details about the eddy footprint area (Figure 1a), such as mean elevation, soil textural types, terrain, and vegetation characteristics, are presented in Table 1 as reported by (Flores, 2021; Marquez, 2023; Mayo, 2023). Regarding vegetation, Flores (2021) reported a 1:4 ratio of Mesquite to Creosote cover in an area adjacent to the eddy tower studied in this research (Table 1). Flores (2021) mapped the areal coverage of mesquite, creosote, and bare soil in a 28-hectare Chihuahuan Desert site using high-resolution drone imagery and supervised classification in ArcMap. By flying in both winter and summer, Flores (2021) distinguished species based on seasonal foliage differences. The imagery was processed into orthomosaics, and a Maximum Likelihood Classification was applied using carefully selected training samples. Accuracy assessments showed over 90% accuracy, with creosote covering 20% of the area, mesquite 5%, and the rest mainly bare ground.

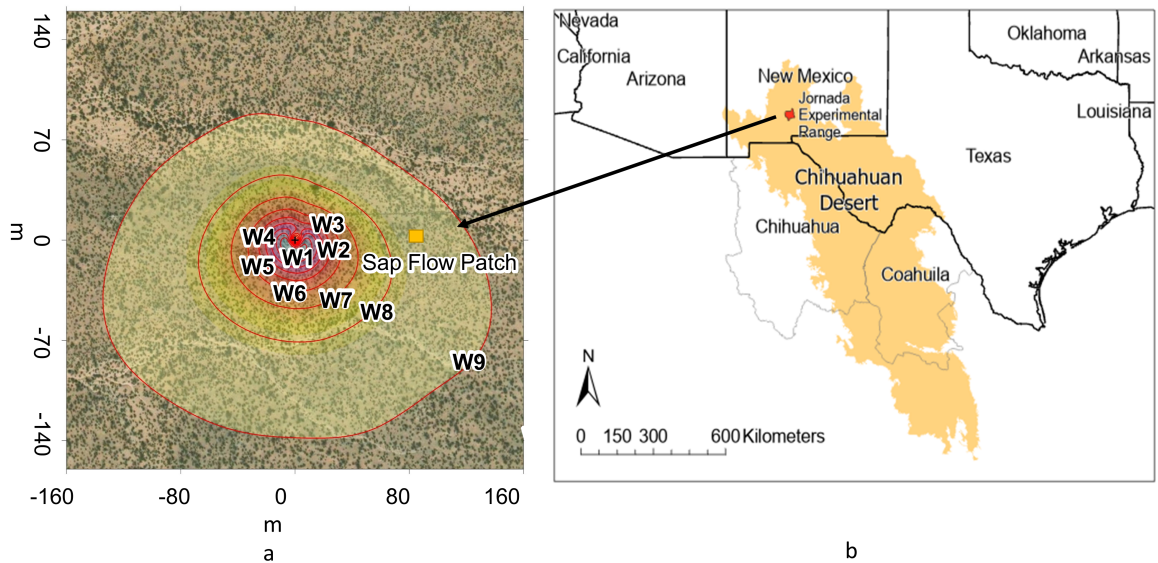

Figure 1: (a) Ameriflux Bajada (US-Jo1) eddy covariance flux footprint computed from the climatological approximation by Kljun et al. (2004) on an RGB raster image from the growing season (summer) of 2022. Contour lines from the most inward ($w_1$) to the most outward ($w_9$) represent the boundary of the footprint contributing areas successively. The outermost red contour line of the source area represents the region that contributes to the ET flux 10% of the time ($w_9$). Successive inward contour lines represent increments of 10% contribution to the ET total flux from the eddy covariance source area up to 90%, with the center of the tower representing the 100% flux contribution. This footprint source area delineates where the measured fluxes originate, with nearer zones contributing more strongly. The image also includes the location of the eight sapflow sensors (yellow box). (b) Location of the Chihuahuan Desert and Jornada Experimental Range (JER) within the southwestern U.S.

## 2.2. Micrometeorological Measurements

This study's data analysis period is from June 1st to September 30th of 2022. At the US-Jo1 site, precipitation (P), evapotranspiration (ET), and other micrometeorological and energy flux variables are continuously measured with quality control assurance protocols (Darrouzet-Nardi et al., 2023; Tweedie, 2024). Precipitation is measured using a Texas Electronics Campbell Scientific tipping bucket rain gauge (TE525-L15-PT), and latent heat flux ($\lambda E$) is monitored with an open-path gas analyzer (LI-7500 LICOR) mounted atop an eddy covariance tower. The sensors are positioned at approximately five meters (except for the rain gauge at 1.2 m) above the ground surface, while the dominant vegetation, primarily creosote bush and honey mesquite, generally ranges from 1 to 2 meters in height, ensuring proper sensor placement above the roughness sublayer. The ET rate is calculated by converting energy flux $\lambda E$ (W.m$^{-2}$) to equivalent water flux (mm.day$^{-1}$) using standard latent heat of vaporization and water density at measured temperatures. All micrometeorological data are aggregated at daily intervals for this analysis.

Table 1: US-Jo1 eddy footprint areal characteristics including terrain, soil textural types, and vegetation (Flores, 2021; Marquez, 2023; Mayo, 2023). The bare soil, Creosote and Mesquite areal coverages were taken from Flores (2021).

| Site Characteristic | Values |
| --- | --- |
| Elevation Range (m) | Min=1,376 m, Max=1,443 m |
| Bare Soil Coverage | 70% |
| Creosote Coverage | 20% |
| Mesquite Coverage | 5% |
| Man-made struct. & shadows | 5% |
| Average Creosote Height | $< 1.5\,\mathrm{m}$ |
| Average Mesquite Height | $< 2\,\mathrm{m}$ |
| Soil Textural Types | Sandy Loam & Silt Loam |
| Caliche Layer Depth | Variable from 0 cm to 60 cm |

*2.3. Plant Sapflux Measurements and Transpiration Values*

A sapflow network of eight sensors was installed on May 28$^{\text{th}}$, 2022, to support this study. Previous research has demonstrated the successful application of these sensors, even in arid and semi-arid environments (Smith and Allen, 1996; Senock and Ham, 1993; Kjelgaard et al., 1997). Eight EXO sap flow (SGEX) devices manufactured by Dynamax, Inc. were deployed on four Mesquite and four Creosote shrubs within a 10 m × 10 m sub-area located approximately 85 m east of the US-Jo1 eddy covariance tower (see Figure 1 and Figure 2). Although not placed at the geometric center of the EC footprint, the selected location lies well within the 80% - 90% flux contribution zone and was chosen based on field validation to ensure it captured the representative species composition, structure, and size diversity of vegetation within the larger EC footprint. This careful placement, combined with UAS-based vegetation mapping and scaling techniques, allows for confident extrapolation of the sapflow measurements to the footprint scale, ensuring that the data are spatially and functionally relevant to the EC measurements.

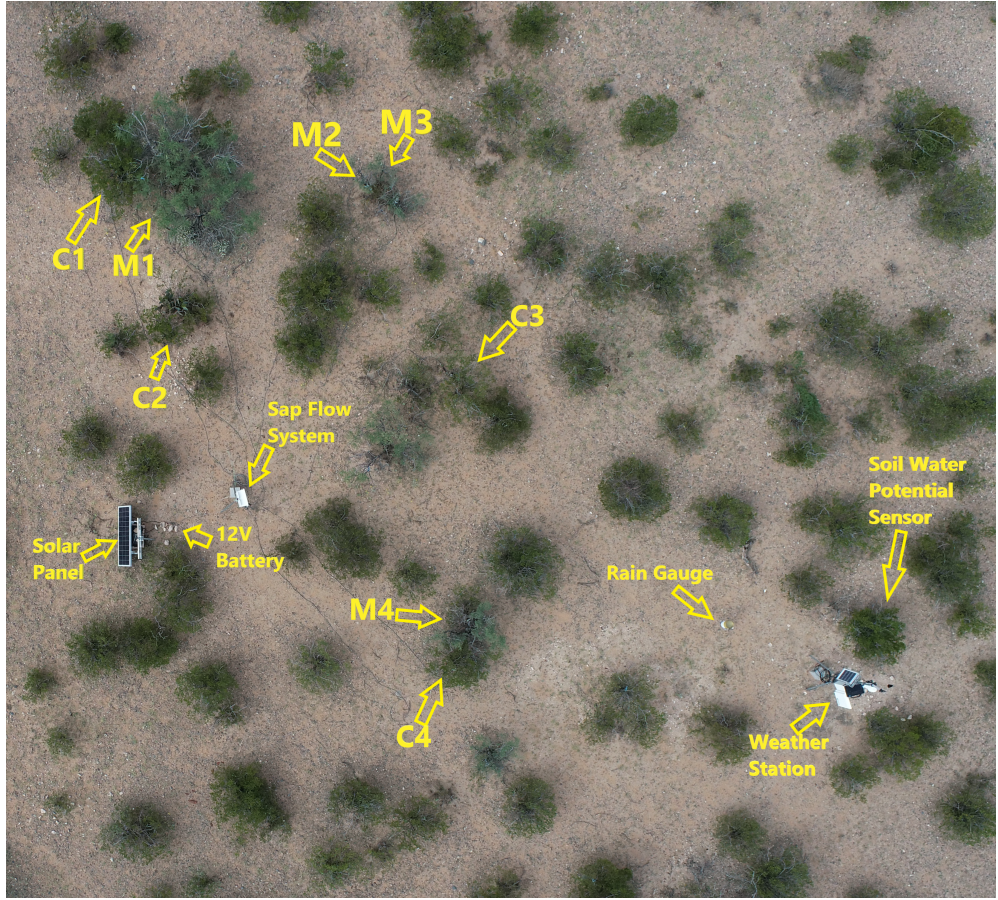

Figure 2: Sapflow patch area within the US-Jo1 eddy tower footprint. The aerial photogrammetry was accomplished using a small UAS 100 m above ground level (AGL). M1 through M4 are the four Mesquite, and C1 through C4 are the four Creosote individuals selected for sapflow monitoring. The picture also shows the location of solar panels, 12V deep cycle batteries, rain gauge, weather station, sapflow logger system, and soil moisture sensors. The centroid of this sapflow patch is located at coordinates 32.5820° North and -106.6350° West and 1,388 m above sea level, about 85 m east of the US-Jo1 eddy tower.

Sap sensor installations were performed on plant individuals of different sizes and aimed to represent the observed range of species within the whole eddy footprint (Figure 2). Figure 3 illustrates the typical sensor installation on a study branch (b) in a Mesquite individual at an arbitrary measurement height (H). The selected height, which ranged from 1 to 1.25 m above ground, was determined individually for each tree based on sap sensor manual recommendations to ensure a minimum branch diameter (i.e. 8 mm) and prevent malfunction. In species like Mesquite and especially Creosote, identifying a single main stem is often not feasible due to their multi-stemmed growth forms, so branches were selected based on accessibility and diameter suitability rather than strict hierarchical position. Installations required careful procedures as indicated by the system user manual. The selected stems were first cleared out with a pointed knife to prevent interference during the installation. Then, sandpaper was used to remove the stem's decaying bark to improve the sensors' contact with the cambium layer. Stems were then wiped with a paper towel and sprayed with canola oil to prevent sensors from adhering. The next stage was to wrap and stretch the double velcro to secure the heater stick's upper and lower thermocouple sections. Soft Gore-Tex material was used to prevent rainfall from penetrating the stem. The next layer of protection consisted of an insulating ring and reflective insulation material to prevent solar radiation from damaging stems (Yue et al., 2008). A data logger and a 12V, 100 Ah deep cycle battery were also installed to maintain the system continuously powered, along with a 75-85 Watt solar panel to provide energy to the system under sunlight. After installation, sensors were set to record heat fluxes at 1-min intervals ($\Delta t$) and then averaged and stored at 30-minute time steps on a Campbell Scientific CR1000X datalogger. Heat fluxes measured by the system are then converted into water flux (F in g) using equation (1) as suggested by Sakuratani (1981); Baker and Van Bavel (1987).

$$F = \frac{P_{\text{in}} - Q_v - Q_r}{c_p \cdot \Delta T} \qquad (1)$$

Where:

$F$ = Total mass of sapflow transport (g) across the measuring branch $b$ during time interval $\Delta t$.

$P_{\text{in}}$ = Power input to the stem from the heater (W).

$Q_v$ = The vertical or axial heat conduction through the stem (W).

$Q_r$ = Rate of heat transfer through radiation (W).

$c_p$ = Specific heat of water (J/g $\cdot$ °C).

$\Delta T$ = Temperature increase of sap (°C).

During the measurement period, the M4 and C4 sapflow sensors malfunctioned due

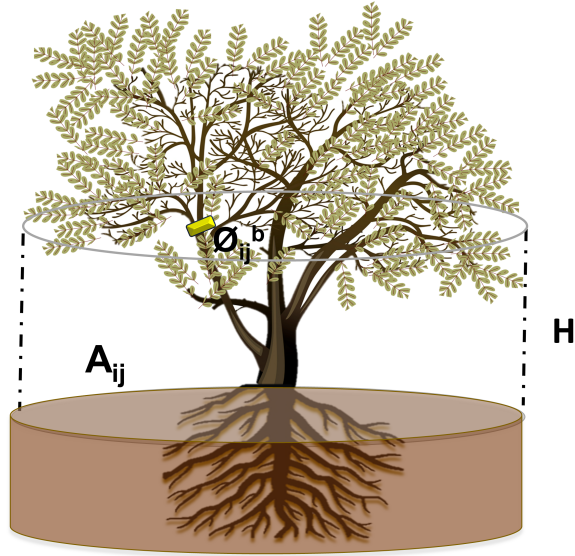

Figure 3: Schematic of a typical Mesquite bush sensor installation at branch "b" of diameter $\phi_{ij}$ ($\phi_{ij}^b$) and other branches at the same height H. This H varies for each plant but roughly ranges from 1 m to 1.25 m above ground. $N_{ij}$ is the total number of branches of different diameters $\phi$ at height H. i is the tree type (i.e., M or C), and j is the tree number (e.g., 1,2,3 and 4). Installations at all Mesquite and Creosote individuals mimic this description

to extremely dry conditions, causing abnormally hot and prolonged periods without data. However, the other plant individuals recorded data correctly. Consequently, the malfunctioning sensors (M4 and C4) were excluded from the analysis. This type of sensor malfunctioning has been reported in previous studies during extremely dry conditions (Cavanaugh et al., 2011).

Sap flux sensors of this type (SGEX) measure flow across the branch cross-section. To estimate whole-plant water use, we assume that sap flux is approximately uniform across all functional branches of a given individual at a given time (t) (Vandegehuchte and Steppe, 2013), a standard approach in multi-stemmed species like Mesquite and Creosote where whole-stem measurements are infeasible. Thus, equation (2) was derived to compute $T_{ij}^\bullet(t)$ as the daily total $T$ (g/d) of vegetation individual type $i$ (e.g., Mesquite, M or Cresote C) and number $j$ (e.g., 1,2,3,4). Note that the next set of equations (2 through 8) provides a time series of values for each time step t at daily temporal resolution.

$$T_{ij}^\bullet(t) = \frac{T_{ij}^b(t)\,\overline{\phi}_{ij} N_{ij}}{\phi_{ij}^b} \qquad (2)$$

Where,

$T_{ij}^b(t)$ = Measured daily transpiration (g/d) along branch $b$ of plant $ij$.

$\overline{\phi}_{ij}$ = Average branch diameter (mm) at sensor height, H of plant $ij$.

$N_{ij}$ = Number of branches at sensor height, H of plant $ij$.

$\phi_{ij}^b$ = Diameter (mm) of branch $b$ of plant $ij$ where the sapflow sensor is installed.

A precision caliper was used to determine $\phi_{ij}^b$, and all other branch diameters $\phi_{ij}$ across the measurement height H up to $N_{ij}$. With these data, a single $\overline{\phi}_{ij}$ per monitored plant was estimated. With $T_{ij}^{\bullet}(t)$ values obtained (in g/d), a subsequent relationship was derived to express this water mass flux as a water depth rate ($T_{ij}(t)$ in mm/d) dividing by the density of water ($\rho$), the typical Leaf Area Index ($LAI_{ij}$) of Mesquite and Creosote and the horizontally projected ground surface area ($A_{ij}$) that each tree occupies (Havstad et al. (2006) (equation 3).

$$T_{ij}(t) = \frac{T_{ij}^{\bullet}(t)}{\rho A_{ij} LAI_{ij}} \qquad (3)$$

Where:

$T_{ij}^{\bullet}(t)$ = Daily transpiration (g/d) of plant $ij$.

$\rho$ = Density of water (0.001 g/mm).

$A_{ij}$ = Horizontal ground projection in mm$^2$ of plant $ij$

$LAI_{ij}$ = Leaf Area Index of plant $ij$

By dividing the sap flow rate by the ground surface area and LAI, we adjust for both spatial plant coverage and foliage density. This yields a transpiration estimate that is more representative of ecosystem-scale water flux than unadjusted sap flow per branch area, which neglects heterogeneity in canopy structure. $A_{ij}$ was measured using low-altitude UAS imagery of each of the eight studied individuals. Since this study did not directly measure leaf area, values were obtained from results reported in the literature described below. According to several consulted studies (Ehleringer et al., 2000; Schwinning et al., 2004; Huxman et al., 2005) LAI of Chihuahuan desert Mesquite and Creosote individuals can vary depending on several factors, including the specific species, age, and environmental conditions. In general, the LAI for Mesquite trees ranges from about 0.9 to 3.9, and Creosote's between 0.2 and 1, including different climatic zones and geographical regions (Ansley et al., 1992; Chopping et al., 2003; Romig et al., 2006; Hamerlynck et al., 2011; Kim et al., 2017; Kirillov et al., 2023). Both species, in the Chihuahuan region, but particularly Mesquite, undergo leaf growth during the summer monsoon season due to higher temperatures and more water availability from the summer rains. Literature estimates for the Jornada Experimental Range provide with LAI values ranging from 0.7 to 3.5 for Mesquite and between 0.3 to 0.8 for Creosote (Gibbens et al., 1996; Havstad et al., 2000; French et al., 2008). Since equation (3) needs a single value for both M and C, $LAI_{Mj}$=2.5 and $LAI_{Cj}$=0.8 were selected to represent the typical conditions of a relatively wet summer in the Chihuahuan desert. 2022 with 168 mm during JJAS represented 30–40% above average summer rainfall. These representative values offer a practical compromise between the observed variability and the need for operational simplicity. While using single-value estimates simplifies analysis, it inherently overlooks the natural variability in LAI—ranging from 0.7 to 3.5 for Mesquite and 0.3 to 0.8 for Creosote bush. To address this, the impact of adopting fixed LAI values rather than full variability ranges will be explicitly assessed through uncertainty envelopes in the final results of this study.

*2.4. Individual Plant to Footprint Transpiration Rates*

The areal extrapolation of the individual plant transpiration estimations ($T_{ij}(t)$) via the sapflow network to the eddy tower footprint ($T_{uv}(t)$), where $u$ is either Mesquite or Creosote and $v$ is the tree number within the footprint, was conducted by understanding: (A) the inter-plant variability of the $T_{ij}^b(t)/\phi_{ij}^b$ term across the days, (B) the density function of $\phi_{ij}$ at each sensor level H for each of the eight sampled plants that allow to come up with a reasonable estimate of $\overline{\phi}_{ij}$, (C) The plant's horizontal projected areas $A_{uv}$ that can be estimated via UAS remote sensing, and (D) A biometric relation between $N_{uv}$ and $A_{uv}$.

(A) and (B) were determined by combining sapflow measurements with a manual inventory of branch density and diameter at the sensor height (H ≈ 1–1.25 m) for each of the eight monitored individuals. For each plant, all live branches intersecting height H were counted and measured to obtain the biometric parameters required in equations (2) and (3). To resolve (C) and (D), a small UAS with a multispectral Micasense RedEdge camera flying 120 m above the ground during the growing season resulted in 3 cm pixel resolution 5-band (i.e., RGB, Red edge, and near-infrared) images of the US-Jo1 eddy covariance footprint. The UAS mission collected 125 images with front and side overlaps of 65% and 70%, respectively. The images were then mosaicked and geo-rectified using Agisoft Metashape. The resulting composite image was classified using a two-step approach. First, a pre-trained Segment Anything Model Kirillov et al. (2023) was fine-tuned in a one-shot manner on the RGB image using a set of 50 (30 C and 20 M) manually labeled samples to segment polygons corresponding to individual vegetation units. Next, the NDVI image was used to classify each segmented individual as either Mesquite or Creosote, based on their distinct spectral signatures and typical sizes. Separate training and testing sets were used to assess the accuracy of both the segmentation and the species-level classification. An iterative thresholding process was conducted, where different NDVI ranges were tested and visually compared against known vegetation patterns and field reference data to optimize class separation between Mesquite and Creosote. Since Creosote typically appeared as much smaller vegetation patches compared to the larger Mesquite canopies, a minimum area threshold specific to Creosote was applied to avoid overestimating its coverage. This threshold helped prevent larger polygons, more likely to be Mesquite, from being incorrectly classified as Creosote based solely on their NDVI value. This refinement ensured that both spectral and structural characteristics were considered in the final classification. After refining the NDVI thresholds and applying the polygon area filter, the resulting classification was validated by cross-checking with high-resolution imagery and field-collected reference points. Accuracy metrics, including overall accuracy, precision, and recall, were computed to evaluate the performance of the segmentation and classification using an independent random set of 60 C and 50 M. This refinement process improved the reliability of distinguishing Mesquite and Creosote in the final vegetation map.

Such a classified vegetation image allowed the measurement of $A_{uv}$ as an attribute of each tree individual reflecting their size and other biometric characteristics. Evenly-distributed, manual sampling was conducted to find $N_{uv}$ for a number of representative (i.e. across ages and sizes) individuals within the 100% (i.e. eddy tower) and 70% source contribution contour line (w=3 in Figure 1) as representative of the footprint vegetation to find a relationship between the number of branches ($N_{uv}$) and horizontally projected area ($A_{uv}$). 33 individuals (16 M and 17 C) were used to develop such a relationship.

Equation (4) was applied to each contributive area (w; see Figure 1) within the eddy flux footprint to derive a unified transpiration term for all M (u=1) and C (u=2) individuals in each area. This calculation was performed for each daily time step corresponding to the sapflow measurements collected at the sapflux patch. The ratio $T^b_{uw}(t)/\phi^b_{uw}$ was taken as the mean T flux value per unit branch for all measured M (u=1) or C (u=2) individuals within the sapflow network patch for each time step t (e.g. for M take the mean$[T^b_{Mj}(t)/\phi^b_{Mj}]$ of all three sampled M, one mean for each time step). So this term corresponds to two time series computed from the three independent measurements for M (u=1) or C (u=2) species for each daily time step t using equation (3).

$$T_{uw}(t) = \frac{T^b_{uw}(t)\,\overline{\phi_{uw}}\,\overline{N_{uw}}}{\phi^b_{uw}\rho\overline{A_{uw}}LAI_{uw}} \qquad (4)$$

Where:

$$T_{uw}(t) \quad = \text{Daily T (mm/d) contribution of u individuals between source contours w-1 and w.}$$

$$u \quad = \text{Vegetation type where 1 is M and 2 is C.}$$

$$w \quad = \text{Lower limit of flux source area, i.e } w=1,2,...9$$

$$T^b_{uw}(t)/\phi^b_{uw} \quad = \text{Mean} \left[\frac{T^b_{ij}(t)}{\phi^b_{ij}}\right] \text{ for M (u=i=1) or C (u=i=2).}$$

$$\overline{\phi}_{uw} \quad = \text{Average branch diameter (mm) at sensor height H of plant type } u.$$

$$\overline{N}_{uw} \quad = \text{Average number of branches at height H for plant species } u.$$

$$\rho \quad = \text{Density of water (0.001 g/mm}^3)$$

$$\overline{A}_{uw} \quad = \text{Average horizontal projection of ground covered area of plant species } u \text{ in mm}^2$$

$$LAI_u \quad = \text{Mean Leaf Area Index for species u.}$$

With all $T_{uw}(t)$ estimations, a weighted average is computed following equation (5) to find out the total transpiration contribution of plant species (u) within each eddy source area (w).

$$T_u(t) = \sum_{w=1}^{n_w} F_w F_{uw} T_{uw}(t) \qquad (5)$$

Where,

$$T_u(t) = \text{Daily transpiration (mm/d) of species u within the footprint.}$$

$$w = \text{Footprint source contour counter, see Figure 1.}$$

$$n_w = \text{Total number of representative source areas. } n_w=9 \text{ if all source areas are considered.}$$

$$F_w = \text{Contributing ET flux fraction of source area w.}$$

$$F_{uw} = \text{Fraction of ground area occupied by all u individuals between contours w-1 and w.}$$

$$u = \text{Tree type. u=1 is M and u=2 is C.}$$

The variable $F_w$ represents the cumulative fraction of the total flux measured by the eddy covariance system that originates from the surface area enclosed within each contour line, from $w_1$ (closest) through $w_9$ (farthest). And the variable $F_{uw}$ accounts for heterogeneities in vegetation type and density distribution within each eddy covariance source area. Nonetheless, if vegetation-covered areas exhibit similar areal coverage proportions across most fractional source areas, an approximation can be made to, for example, w can be set to 1 (100% to 90% source contribution) or allowed to vary between 1 and 2 (100% to 80% source contribution), assuming that the source area adequately represents the vegetation distribution of the remaining footprint. This has to be verified via the probability functions of the areal coverage fraction of each vegetation type. This simplification can reduce computational complexity while maintaining accuracy in modeling the spatial variability of vegetation. Finally, the total plant transpiration per time step (t) within the representative eddy footprint is computed as equation (6):

$$T(t) = \sum_{u=1}^{n_u} T_u(t) \qquad (6)$$

## 3. Results

*3.1. Biometric Measurements of Sapflow Patch Plants*

Figure 4 (a and b) shows the fitted frequency distributions of plant branch diameters ($\phi_{ij}$) at measurement height H for the four Mesquite (M) and four Creosote (C) individuals with sapflow sensors. The mean values ($\overline{\phi}_{ij}$) of these diameters are provided in Table 2. Figure 4 indicates that Mesquite individuals have branch diameters ranging from 6.8 mm to 18.3 mm, while Creosote individuals range from 5.1 mm to 12.3 mm. Among the Mesquite plants, M3 has the thinnest branches, whereas M1, M2, and M4 have thicker branches with narrower distribution spreads compared to M3. On the other hand, C individuals present slightly smaller mean diameter values (compared to all M), but their distributions appear more similar in terms of data dispersion around the mean.

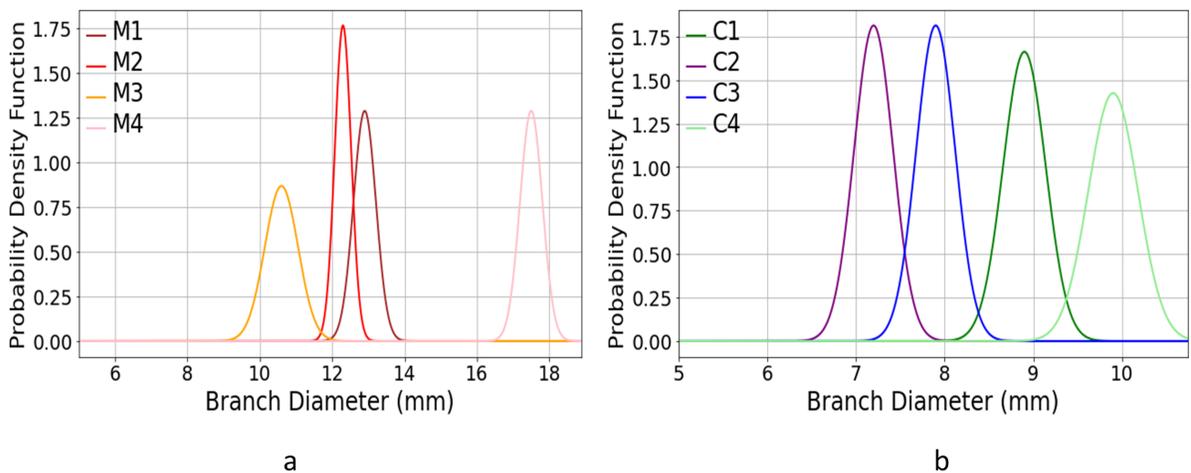

Figure 4: Probability density functions of the branch diameters, $\phi_{ij}$, at the sensor measurement height (H) for the (a) four Mesquite and (b) four Creosote observed individuals

Table 2 presents the values of plant biometric features related to equations (2) and (3) for the sap flow-installed M and C plant branches. The branch diameter ranges with sensor installation ($\phi_{ij}^b$) are 14.6 mm to 18.2 mm for M and 9.2 to 11.4 mm for C. These values are slightly larger than the average branch diameter values $\overline{\phi}_{ij}$ (Table 2, column 3) at height H (Figure 3) as suggested by the sensor manual to avoid overheating.

Table 2: Plant biometric features for the sapflow-installed sensors at $ij$ plants. M means Mesquite, and C means Creosote (see Figure 2 for further spatial details). $\phi_{ij}^b$ is the monitored plant branch diameter, $\overline{\phi}_{ij}$ is the mean branch diameter, $A_{ij}$ is the canopy projected surface area and $N_{ij}$ is the counted number of branches of species $i$ and individual $j$.

| Plant (ij) | $\phi_{ij}^b$ (mm) | $\overline{\phi}_{ij}$ (mm) | $A_{ij}$ (m$^2$) | $N_{ij}$ |
|---|---|---|---|---|
| M1 | 16.2 | 12.9 | 2.8 | 20 |
| M2 | 14.6 | 12.3 | 0.94 | 16 |
| M3 | 17.2 | 10.6 | 0.62 | 14 |
| M4 | 18.2 | 17.5 | 2.15 | 17 |
| C1 | 9.5 | 8.9 | 2 | 28 |
| C2 | 9.2 | 7.8 | 1.16 | 25 |
| C3 | 10.3 | 7.9 | 1.92 | 22 |
| C4 | 11.4 | 9.9 | 1.27 | 20 |

The ground projected areas $A_{ij}$ (Table 2) appear similar across M and C individuals and perhaps related to age (Niklas, 1994; Waring and Schlesinger, 1985) with values ranging from 0.62 m$^2$ to 2.8 m$^2$ for M and 1.16 m$^2$ to 2.0 m$^2$ for C. The largest areas are occupied by M1 (2.8 m$^2$), M4 (2.15 m$^2$), C1 (2.0 m$^2$) and C3 (1.92 m$^2$). Finally, the number of branches at height H ($N_{ij}$) illustrates values ranging from 14 to 20 for M and 20 to 28 for C. Therefore, although C individuals tend to have a smaller diameter, the number of branches tends to be higher on average at measurement height H compared to M individuals.

*3.2. Transpiration Rate Per Unit Branch Diameter*

Figure 5 (a and b) illustrates a time series with precipitation (P; mm/d) and the ratio $T_{ij}^b/\phi_{ij}^b$ (g/mm.d; see equation 2) for each of the six observed trees (recall that M4 and C4 sensors suffered malfunctioning). Overall, $T_{ij}^b/\phi_{ij}^b$ responds to precipitation inputs in the subsequent days after the water inputs (note the sustained increase after the late August Monsoonal events in Figures 5 (and b)). Figure 5(a) indicates that the three M individuals present similar $T_{ij}^b/\phi_{ij}^b$ with M1 and M3 more alike. Throughout the entire summer season, M2 displayed lower $T_{ij}^b/\phi_{ij}^b$ values in comparison to M1 and M3, except by some days in mid June and mid July when peaks were observed in M2. Similarly, C individuals (Figure 5(b)) present consistent (but lower than M) $T_{ij}^b/\phi_{ij}^b$ values across the summer with increases after the mid-August Monsoonal events. Overall, C2 and C1 present higher values than C3 over the measurement period.

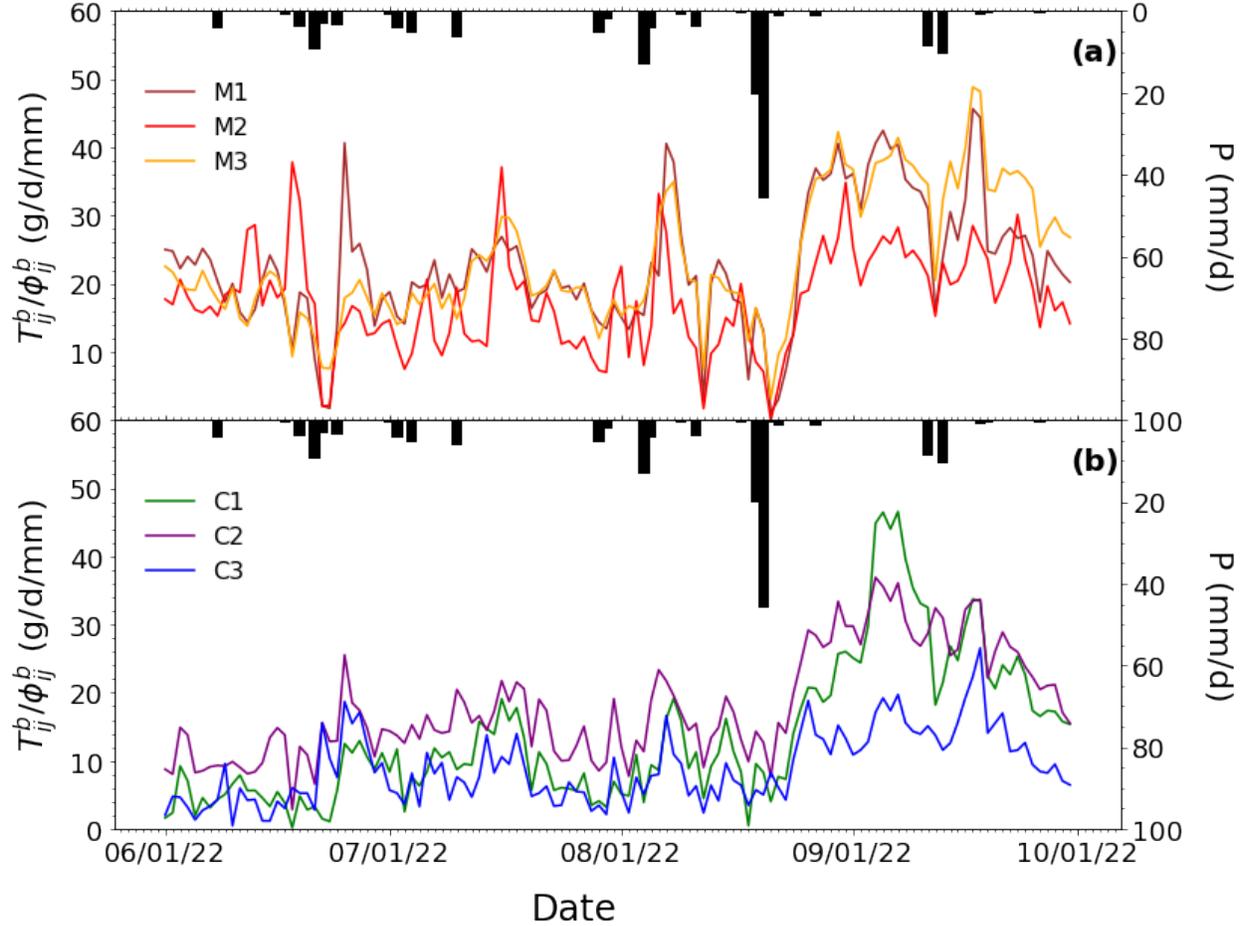

Figure 5: Daily time series of plant-level transpiration ($T_{ij}^b/\phi_{ij}^b$) (g/mm.d) on left y-axis and precipitation (P) (mm/d) on the right y-axis based on the sapflow-observed for individual (a) Mesquite M 1-3 and (b) Creosote C 1-3 shrubs during study period.

*3.3. Transpiration Rate Per Ground Covered Leaf Area*

Figures 6 (a and b) show the behavior of plant transpiration $T_{ij}$ (mm/d) for the sap flow-observed M (Figure 6(a)) and C (Figure 6(b)) individuals. The time series were obtained after applying equation (3) with the values from Figure 5 and Table 2. Overall, the average daily T rates for the different individuals during the four months (JJAS) of 2022 are $\overline{T_{M1}}$=1.98 mm/d, $\overline{T_{M2}}$=4.17 mm/d, $\overline{T_{M3}}$= 2.37 mm/d, $\overline{T_{C1}}$=1.93 mm/d, $\overline{T_{C2}}$=2.32 mm/d and $\overline{T_{C3}}$=1.09 mm/d. Average transpiration values for the three M ($\overline{T_{Mj}}$=2.84 mm/d) plants appear to be 1.6 times higher (on average) than those of C plants ($\overline{T_{Cj}}$=1.78 mm/d). This ratio appears to be slightly higher during the June to August period (drier conditions) when $\overline{T_{Mj}}$=2.5 mm/d while $\overline{T_{Cj}}$= 1.37 mm/d ($\overline{T_{Mj}}/\overline{T_{Cj}}$=1.8). On the other hand, when the strongest monsoon precipitation events appear in September, $\overline{T_{M2}}$=5.56 mm/d, $\overline{T_{M3}}$= 3.544 mm/d while $\overline{T_{M1}}$=2.45 mm/d. For Creosote, the September storm showers resulted in a more significant increase in transpiration to $\overline{T_{C2}}$=3.55 mm/d $\overline{T_{C3}}$=1.69 mm/d and $\overline{T_{C1}}$= 3.88 mm/d. Across the month of September (rainier period) $\overline{T_{Mj}}/\overline{T_{Cj}}$=1.3, which means that overall after precipitation occurs, Creosote bush increases their transpiration rates more significantly than Mesquite. C1, which has the highest number of branches across the C individuals ($N_{C1}$=28), appears to have the highest T values across the C individuals during the drier and storm periods, sometimes approaching the M1 transpiration rates. The $T_{M1}$ rates are the lowest of the three M individuals.

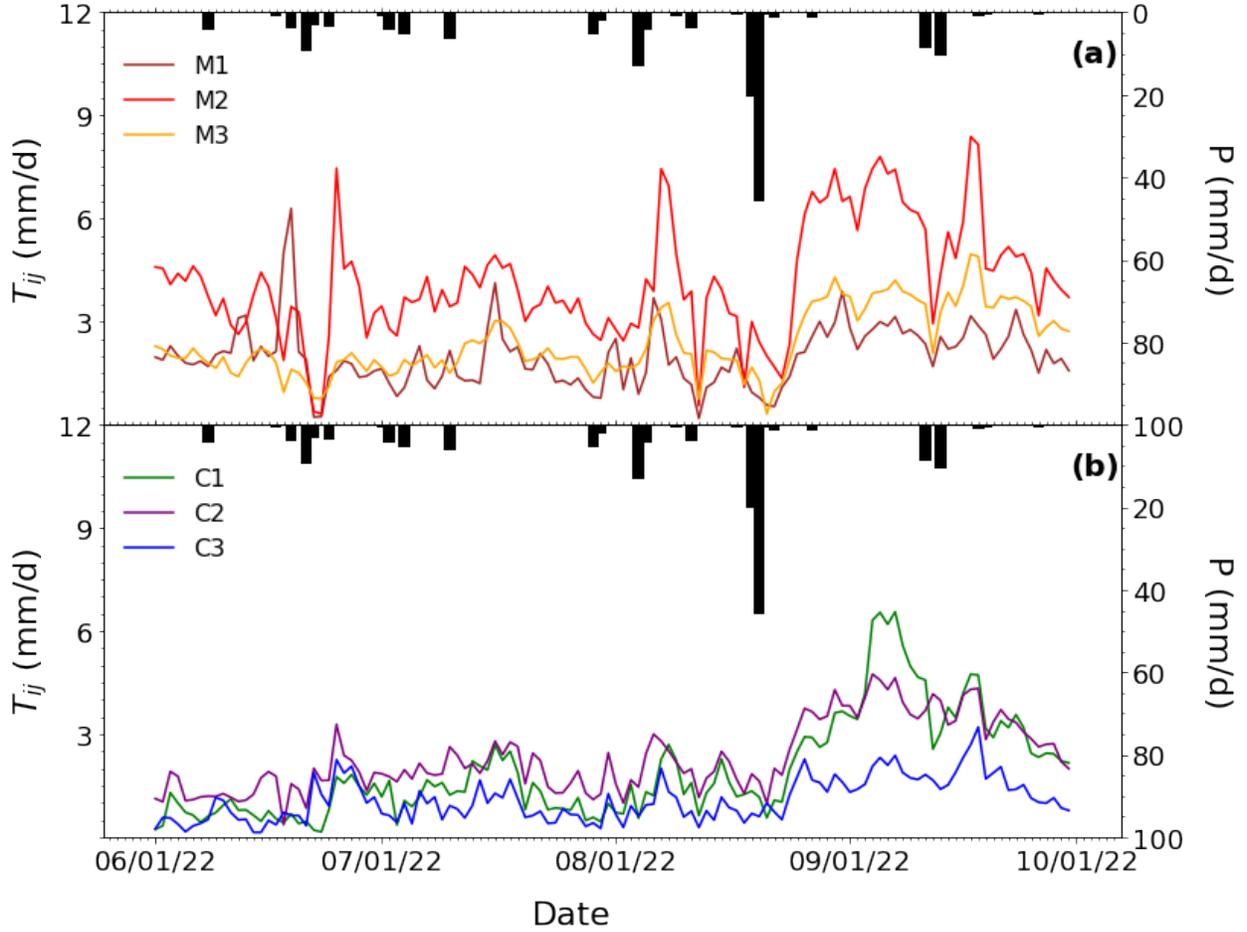

Figure 6: Daily time series of T$_{ij}$ (mm/d) and precipitation (mm/d) for the sapflow-observed M and C shrubs during the study period

*3.4. Footprint-Scale Transpiration Rates*

The areal estimation of total footprint transpiration used the UAS-derived multispectral image over the entire eddy covariance flux footprint (up to the 10% contribution area or $w_9$; see Figure 1). A one-shot fine-tuning of the Segment Anything Model (SAM) was performed using ArcGIS Pro v3.4's object detection workflow, with model parameters set to a padding of 64, batch size of 64, a learning rate of 1e-4 and ResNet34 as a backbone. The model was trained for 1024 iterations and subsequently used to detect and extract all vegetation polygon objects across the image.

Following segmentation, a rule-based object classification was applied based on the relative dominance of NDVI pixel values within each canopy segment. Although Mesquite generally exhibits higher overall vegetation activity, not all its pixels consistently show high NDVI values—some areas may appear dormant or inactive. Conversely, Creosote tends to display lower NDVI values, though isolated pixels may show increased greenness. To establish an appropriate NDVI threshold for distinguishing the two species, a visual inspection of their typical NDVI profiles was conducted. This analysis confirmed that Mesquite consistently exhibited higher NDVI values, justifying the use of a single threshold for classification.

Each segmented object was then evaluated by counting the number of pixels above and below the NDVI threshold of 0.25. An object was classified as Creosote if it had a majority of pixels with NDVI < 0.25 (up to 75% coverage), along with a minority of pixels with NDVI $\geq$ 0.25 not exceeding 8% coverage. In contrast, an object was classified as Mesquite if it contained at least 15% of pixels with NDVI $\geq$ 0.25 and had a total area of at least 1500 pixels (camera resolution was 3 cm/pixel). These classification rules were refined through iterative testing to align with field observations, where Mesquite typically forms larger, denser canopies, and Creosote occurs as smaller, more scattered shrubs.

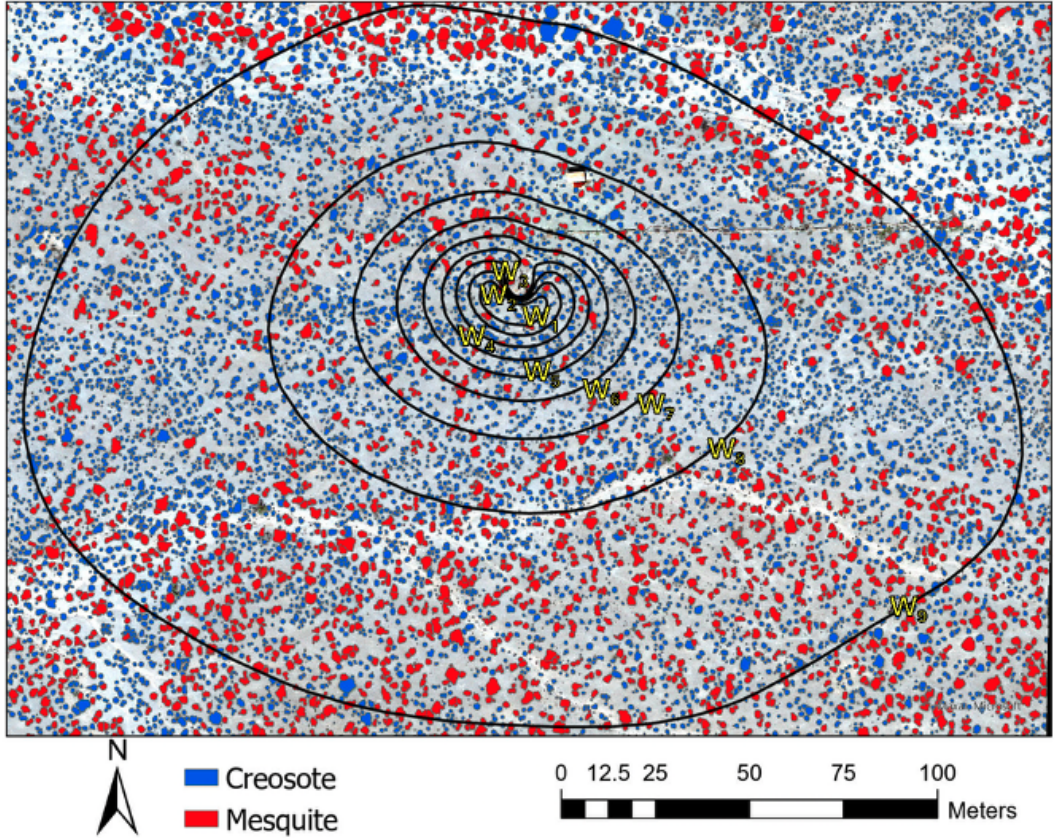

Figure 7: Deep learning based, supervised segmentation and classification of the UAS-obtained 5-band image at 3 cm pixel resolution within the eddy footprint area of the US-Jo1 eddy covariance tower. The two main vegetation classes shown are Mesquite (red color) and Creosote (blue color). The outermost green contour represents the 10% (percent of the time) vapor flux source area, while subsequent inner contours represent increments of 10% in temporal contribution to total eddy-measured ET.

The final object-based classification map is shown in Figure 7. To evaluate the accuracy of the approach, a validation set consisting of 50 Mesquite and 60 Creosote individuals was randomly selected and assessed. The model achieved an overall classification accuracy of 86.4%, with precision of 78.2%, recall of 78.0%, and F1 score of 78.0%.

Table 3 summarizes some relevant statistics and calculations from Figure 7 regarding M and C individuals per footprint contribution area from $w_1$ to $w_9$. For instance, the total footprint area until $w_9$ is 68,496 m$^2$ (Column 3 in Table 3) and the number of M and C individuals within the footprint (Columns 5 and 11) is $n_M$=1898 and $n_C$=8919, respectively (a ratio M:C of 1:4.7). The fractional areas covered by M and C within each contribution area ($F_{uw}$) vary from 0.11 to 0.45 for M and 0.54 to 0.89 for C and the mean ground covered areas $\overline{A_{uw}}$ vary from 3.05 m² to 4.63 m² for M and 1.15 to 1.36 m² for C. From the total footprint area, M and C cover 12.6% and 17.38%, respectively. Therefore, bare soil accounts for 70% of the surface footprint source area. These values are close to the ones reported in Table 1, except for M that shows a larger fraction as the referenced articles focused in the proximity of the eddy tower where C clearly dominates (See Table 3, columns 7 and 13 for $F_{Mw}$ and $F_{Cw}$).

Figure 8 illustrates a proportional pattern between $N_{uv}$ and $A_{uv}$ that was approximated by near-linear relationships with a coefficient of determination ($R^2$) above 0.85. Overall, Cresote plants show smaller covered areas (0.5 m² ≤ $A_{Cv}$ ≤ 1.3 m²) than Mesquite (1.2 m² ≤ $A_{Mv}$ ≤ 3 m²) while the number of branches is slightly higher for C (18 ≤ $N_{Cv}$ ≤ 27) than M (15 ≤ $N_{Cv}$ ≤ 24). Furthermore, the slope of the relationship between these two variables ($N_{uv}$ vs $A_{uv}$) is higher for Creosote individuals. This relationship was applied to all individuals classified in Figure 7 and the results are shown in columns 10 and 16 of Table 3 (i.e. $N_{Cw}$ and $N_{Mw}$).

Given the relative homogeneity of stem diameters across individuals of the same species, as shown by Table 2 and Figure 4, it was decided that $\overline{\phi}_{uv}$ should be taken as the mean value per species. So, for Mesquite, $\overline{\phi}_{Mv} = 13.6$ mm, while for Creosote $\overline{\phi}_{Cv} = 8.6$ mm. Concerning the $T^b_{ij}/\phi^b_{ij}$ to be applied to each of the *uv* trees, the results shown in Figure 5 support the selection of an average rate for all trees of the same species per time step t, as described in section 2.4.

Based on the inputs described above, $T_{uv}$ was calculated for each footprint contribution zone from $w_1$ to $w_9$, without simplification, to account for the distinct spatial distribution of species across source areas (see Table 3). The resulting values, obtained using equations (4) and (5), are presented in Figure 9.

Besides the time series of daily T (mm/d) values, Figure 9 also illustrates total daily P (mm/d) and the eddy-covariance measured ET (mm/d) values with commonly accepted error envelopes of 15% above and below the measurements (Novick et al., 2009; Denager et al., 2020). The total P during the JJAS summer period of 2022 was 168.1 mm, while the total ET was 140.2 mm and T= 70.01 mm. Therefore, for this period, ET/P= 0.83, T/ET=0.50, and E/ET= 0.50. Figure 9 also illustrates T values with an error envelope given by the uncertainty in $T_{ij}$ computed as the mean absolute error of the variability introduced from Figure 6 (intra-individual variability) and the range of summer typical LAI values ($0.7 \leq \text{LAI}_{Mj} \leq 3.5$ and $0.3 \leq \text{LAI}_{Cj} \leq 0.8$.

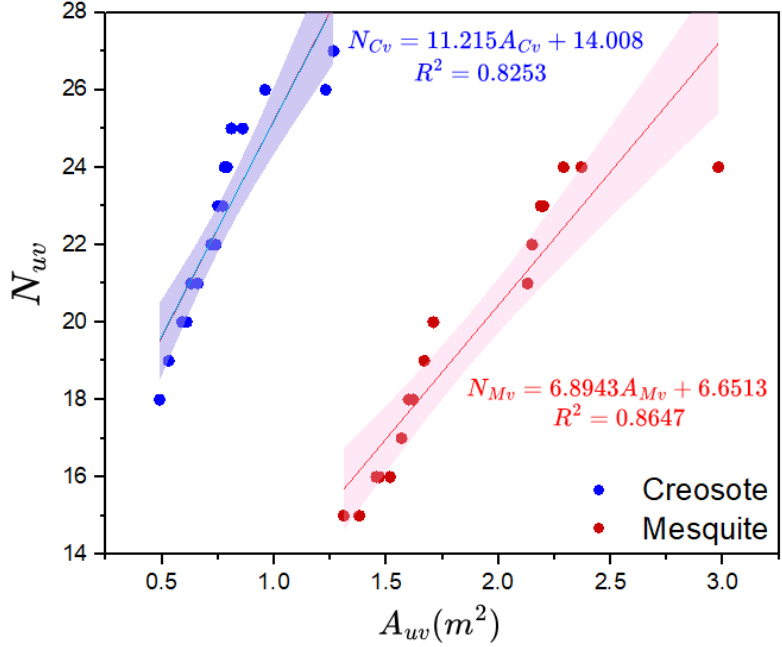

Figure 8: $N_{uv}$ as a function of $A_{uv}$ for 17 M and 17 C randomly selected individuals within the 70% ET contribution region. Linear regression equations are suggested with 20% uncertainty envelopes.

Figure 9 shows that transpiration (T) is almost always below evapotranspiration (ET) and its error envelopes. It can be observed that the immediate response of ET to precipitation (P) events primarily comes from evaporation (E), while T shows delayed responses due to vegetation's slower access to water in the root zone, which can take several hours to days after the main storm events. However, once vegetation accesses root-zone water, T accounts for a more significant portion of ET while E is already experiencing a recession. During June, July, and August, with relatively low precipitation inputs (P=73.6 mm), ET values average 1.05 mm/d, while T averages 0.46 mm/d (T/ET=0.44). After the more intense precipitation events in early September (P=94.5 mm for the entire month), the mean ET values rise to 1.43 mm/d, and T rises to 0.91 mm/d (T/ET=0.63) on average from September $1^{st}$ to September $30^{th}$.

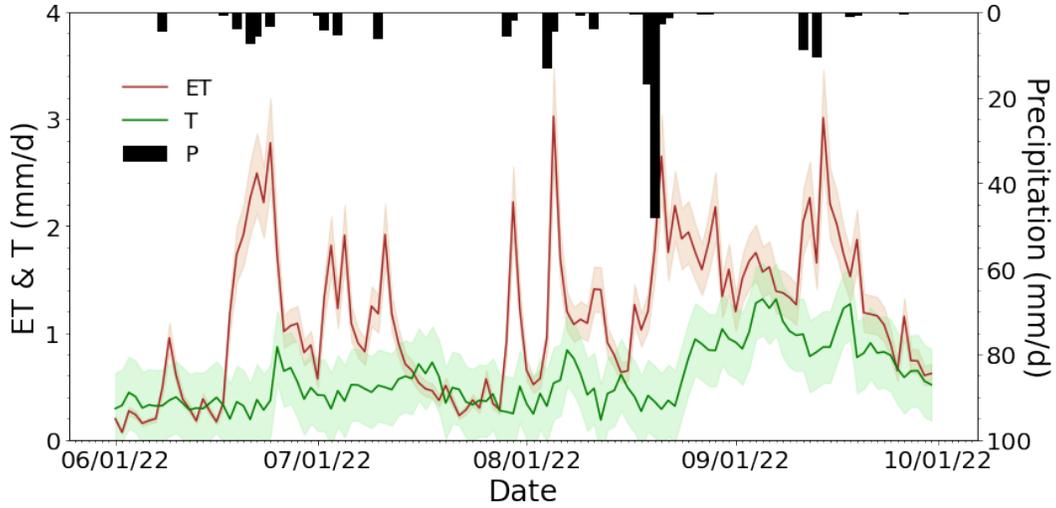

Figure 9: Time series of mean daily precipitation (P, mm/d; black bars), footprint-scale evapotranspiration (ET, mm/d; red line) from the AmeriFlux US-Jo1 eddy covariance system and transpiration (T, mm/d; green line). T was calculated using equations (4), (5) and (6). The shaded red and green envelopes represent the estimated uncertainty for ET and T, respectively, as described in Section 3.4. Note that the right y-axis represents precipitation in reverse scale, consistent with hydrological convention.

Table 3: Plant species count, mean, total and standard deviations of ground covered areas of M and C individuals within each eddy footprint source region from $w_1$ to $w_9$. $A_w$ is the total area enclosed between $w_{i-1}$ and $w_i$ contours and $F_w$ is the contribution fraction of such source area to total ET.

| Source | Cont. (%) | $A_w$ (m$^2$) | $F_w$ | $n_{Mw}$ | $A^T_{Mw}$ (m$^2$) | $F_{Mw}$ | $\overline{A_{Mw}}$ (m$^2$) | $\sigma(A_{Mw})$ (m$^2$) | $N_{Mw}$ | $n_{Cw}$ | $A^T_{Cw}$ (m$^2$) | $F_{Cw}$ | $\overline{A_{Cw}}$ (m$^2$) | $\sigma(A_{Cw})$ (m$^2$) | $N_{Cw}$ |
|---|---|---|---|---|---|---|---|---|---|---|---|---|---|---|---|
| $w_1$ | 90-100 | 132.41 | 0.20 | 0 | 0 | 0 | 0 | 0 | 0 | 24 | 31.11 | 1 | 1.30 | 0.66 | 28.54 |
| $w_2$ | 80-90 | 307.01 | 0.18 | 8 | 32.13 | 0.43 | 4.02 | 1.12 | 34.34 | 33 | 43.28 | 0.57 | 1.31 | 0.87 | 28.72 |
| $w_3$ | 70-80 | 579.44 | 0.16 | 5 | 18.71 | 0.17 | 3.74 | 2.47 | 32.45 | 69 | 88.38 | 0.83 | 1.28 | 1.06 | 28.37244 |
| $w_4$ | 60-70 | 981.54 | 0.13 | 5 | 15.24 | 0.11 | 3.05 | 0.86 | 27.67 | 99 | 127.52 | 0.89 | 1.29 | 1.21 | 28.45 |
| $w_5$ | 50-60 | 1669.55 | 0.11 | 14 | 59.67 | 0.26 | 4.26 | 2.01 | 36.04 | 149 | 171.55 | 0.74 | 1.15 | 0.7557 | 26.92 |
| $w_6$ | 40-50 | 2798.88 | 0.09 | 34 | 134.51 | 0.27 | 3.96 | 1.66 | 33.93 | 281 | 355.63 | 0.73 | 1.27 | 1.3954 | 28.20 |
| $w_7$ | 30-40 | 5156.57 | 0.07 | 57 | 234.43 | 0.27 | 4.11 | 2.41 | 35.01 | 479 | 634.28 | 0.73 | 1.32 | 1.0513 | 28.85861 |
| $w_8$ | 20-30 | 11690.12 | 0.04 | 211 | 860.10 | 0.33 | 4.08 | 2.18 | 34.75 | 1402 | 1786.58 | 0.67 | 1.27 | 1.17 | 28.30 |
| $w_9$ | 10-20 | 45180.05 | 0.02 | 1564 | 7241.74 | 0.45 | 4.63 | 3.15 | 38.57 | 6383 | 8663.93 | 0.54 | 1.36 | 1.45 | 29.23 |

## 4. Limitations

Despite striving to come up with a physically-sound and practical method that incorporates high-resolution UAS imagery to scale measurements up from a smaller vegetation sap flux patch to the eddy covariance footprint region, the method relies on several assumptions that might pose limitations for its further use in other regions, but that can definitely encourage further tests and subsequent studies:

1. A key limitation of this study lies in the restricted number of sapflow sensors deployed, eight in total, due to budgetary and logistical constraints. The installation and continuous operation of high-quality sapflow monitoring systems, including data loggers and power sources (e.g., solar panels), incurs substantial costs, estimated at approximately USD $10,000 for this experiment. While this scale of monitoring may not fully capture the physiological variability among all individuals within the footprint, our sensor network was designed to span a representative range of plant sizes and branch structures. Moreover, this constraint is not unique to our study; similar limitations are reported in other footprint-scale transpiration studies (e.g., (Scott et al., 2006a; Moran et al., 2009; Cavanaugh et al., 2011), which often utilize fewer than ten sensors due to the high cost and complexity of long-term monitoring. These methodological parallels highlight the practical value of our approach: its transferability to other ecosystems facing similar resource limitations. Nevertheless, future studies would benefit from denser sapflow networks, particularly to explore inter- and intra-species transpiration variability more comprehensively and improve uncertainty quantification.

2. Another limitation relates to the choice of branch-mounted rather than stem-mounted sap flow sensors. While stem-based installations can theoretically provide an integrated sap flow estimate per plant, the low and irregular growth forms of Mesquite and especially Creosote bushes in our study area posed substantial physical limitations for proper sensor installation. Moreover, thermal stability, xylem contact, and installation standardization are often more reliably achieved on secondary branches with uniform diameters. Thus, while stem measurements could reduce the need for branch-based extrapolations, our approach allowed for consistent, replicable monitoring aligned with previous studies in similar environments (e.g.,(Smith and Allen, 1996; Hultine et al., 2004; Cavanaugh et al., 2011; Burgess and Dawson, 2007). We do acknowledge the assumptions made in scaling up branch measurements and have explicitly accounted for these in our uncertainty analysis.

3. Total sap flow per plant (g/d) was estimated as proportional to both the number and diameter of branches within each individual. This approach assumes that all live branches at the measurement height (H) transport sap at similar rates per unit cross-sectional area (Berry et al., 2017; Jensen et al., 2016). By multiplying this rate by the number of branches, we approximate the total sap transport per plant. However, a key limitation of this study is the lack of a diameter-based scaling function for sap flow. Due to the small sample size (n = 8) and limited variation in branch diameter, it was not feasible to derive a robust relationship. Future research with broader sampling is needed to develop transferable scaling functions linking sap flow to branch diameter or sapwood area. In addition, while using mean diameter values per species simplifies the scaling of sap flow measurements, this approach does not fully account for inter-branch variability within individuals. As a result, this may contribute to uncertainty in footprint-scale transpiration estimates. Future work with expanded sampling should explore non-linear scaling relationships between sap flow and branch diameter to improve estimation fidelity. Finally, while a linear relationship between canopy area and the number of branches was assumed for scaling purposes (Figure 8), this may not hold for Mesquite individuals with smaller canopies (A $<$ 1.0 m$^2$), which were underrepresented in the sap flow sampling. As a result, the model may underpredict the contribution of smaller individuals to overall transpiration. Given that these individuals comprise a substantial portion of the EC footprint, future work should assess whether a nonlinear or stratified model could better account for their role in total water fluxes.

4. An inherent simplification in our upscaling approach lies in the conversion of sap flow from a mass flux (g d$^1$) to a depth-based rate (mm d$^1$) using Equation 3. This requires dividing by both the ground area covered by the plant and its leaf area index (LAI). Because direct measurement of LAI for each individual plant is impractical in field settings, we rely on representative LAI values derived from limited sampling and species averages. While this simplification allows for tractable estimation of transpiration per unit ground area, it introduces uncertainty in capturing the full variability of canopy density and leaf distribution across the footprint. Nonetheless, this adjustment provides a more spatially and structurally representative estimate of transpiration than raw sap flow scaled only by sapwood area, and helps better reflect ecosystem-scale water use.

5. There is a relationship between the number of branches at the same (hypothetical) sapflow measurement height H and the plant's ground projected area that can be extended to most (if not all) plants within the eddy footprint. The number of branches at a given sap flow measurement height H often indicates a plant's horizontal canopy coverage, which directly influences its transpiration rate. More branches generally correlate with a larger canopy area and, consequently, a higher transpiration rate, assuming other factors are constant (Goulden et al., 1997; Oren et al., 1999; Wilson and Baldocchi, 2001; Katul et al., 2000; Monteith, 1965).

6. Time series of values representing plant transpiration per unit plant diameter per species type are indispensable for accurately accounting for species-specific differences in transpiration within the eddy covariance footprint because these time series data provide critical insights into how various plant species contribute differently to overall water vapor fluxes. Plants of different species exhibit distinct physiological and structural characteristics that influence their transpiration rates, such as variations in leaf area, stomatal density, and hydraulic efficiency (Granier et al., 2000; Dawson and Ehleringer, 1991; Oren et al., 2001; Gough et al., 2008; Law and Waring, 1994). Additionally, using not one but several sapflow sensors across individuals of the same species is fundamental to understanding the physiologic and structural differences among individuals of the same species but of different ages (or sizes) that allow capturing intra-species variability in transpiration rates.

## 5. Discussion

Despite the presented limitations, this paper introduces and develops a methodology for estimating transpiration values across an eddy covariance footprint through the use of an in-footprint sapflow observation patch, the measurement of essential plant biometric features (i.e., typical branch diameter distributions and number of branches), and a low-altitude footprint ortho-photo. The method is applied to a fetch region with dominant Mesquite and Creosote bush species in the Jornada Experimental Range of the Chihuahuan Desert in southern New Mexico, United States. The results are assessed in terms of the total measured ET and analyzed regarding the precipitation inputs and the contribution of each plant species to the total T.

The sap flow measurements across the footprint revealed consistent species-level and seasonal differences in transpiration dynamics, as shown in Figures 5 and 6. Mesquite individuals showed significantly higher sap flow rates than Creosote, particularly following monsoon rainfall events. This trend supports the interpretation that Mesquite's deeper rooting system, higher canopy conductance, and greater leaf area index allow it to respond rapidly to soil moisture pulses (Huxman et al., 2004; Ogle and Reynolds, 2004; West et al., 2007). Creosote, in contrast, exhibited more conservative water-use behavior, with subdued responses to precipitation and lower transpiration overall—traits aligned with its adaptation to chronic aridity and minimal stomatal regulation (Gibbens et al., 1996; Ehleringer et al., 1991). These differences are quantitatively reflected in Table 3, which shows higher average branch counts and cross-sectional area for Mesquite, particularly in central footprint zones, reinforcing its greater contribution to total transpiration. Figures 5 and 6 also illustrate the close coupling between sap flow dynamics and short-term precipitation pulses, a hallmark of opportunistic water-use strategies in arid and semiarid ecosystems (Schwinning et al., 2003; Novick et al., 2016). Footprint-scale results in Figure 9 further support the role of vegetation structure and species composition in shaping transpiration patterns. Sap flow-derived transpiration showed strong temporal alignment with ET from eddy covariance, and the proportion of transpiration to total ET (T/ET) varied markedly across time and space. During dry periods, T/ET remained below 0.3, indicating that soil evaporation dominated water loss; however, following major rainfall events, T/ET often exceeded 0.5 and approached 0.7 in Mesquite-dominated zones such as $w_3$ and $w_5$. These dynamics are consistent with theoretical expectations from the Budyko framework, where transpiration dominates under higher vegetation activity and soil moisture availability (Zhang et al., 2001; Yang et al., 2008). The spatial heterogeneity shown in Figures 7 and 8, combined with Table 3's vegetation structure metrics, underscores the necessity of considering plant functional diversity and distribution in partitioning ET. Although our upscaling relied on simplifying assumptions regarding sap flux uniformity, this approach, when constrained by biometric data and canopy structure, has proven useful for dryland ET disaggregation (Oishi et al., 2008; Stoy et al., 2019). Future studies should aim to integrate remote sensing and sapwood-based allometric scaling to improve the robustness and transferability of these estimates.

To the best of our knowledge, the novelty of this work can be summarized as follows: 1. Previous studies have often focused on understory species in temperate and subtropical shrublands Terminel et al. (2020) or have integrated species-specific data to improve the accuracy of estimating evapotranspiration (ET) at broader spatial scales Ewers et al. (2008). Others have investigated multi-species sap flow patterns in savannas and forests to understand species contributions to water fluxes (Eamus et al., 2013; Wilson et al., 2001; Tang et al., 2004), though typically outside formally defined eddy covariance (EC) footprints. On the other hand, several studies have attempted to partition evapotranspiration (ET) within the eddy covariance (EC) footprint using a variety of approaches. These include integration with remote sensing and modeling Herbst et al. (2008), partitioning latent and sensible heat fluxes from EC data, catchment-scale water balances Humphreys et al. (2003); Jha et al. (2013), and sap flow measurements from nearby transects Cavanaugh et al. (2011). Others have utilized vegetation structural metrics such as leaf area index (LAI), fraction of photosynthetically active radiation (FPAR), and vegetation fractional cover (FVC) to support partitioning (Pu et al., 2020; Zeng et al., 2000). Building on these approaches, a number of recent and widely used methods have advanced ET partitioning. These include the Functional Vegetation Stratification (FVS) method (Scanlon et al., 2019; Skaggs et al., 2018), the underlying Water Use Efficiency (uWUE) method Zhou et al. (2016), and the Transpiration Estimation Algorithm (TEA) Nelson et al. (2020). Other notable techniques include the Pérez-Priego et al. (2018) method, conductance-based partitioning Li et al. (2019), conditional eddy covariance Zahn et al. (2022), and the Eichelmann et al. (2021) approach, which utilize physiological or flux-based partitioning frameworks. The Scott and Biederman (2017) and the recently developed DEPART method Reich et al. (2024) are particularly relevant in dryland ecosystems, addressing the specific challenges of sparse canopy cover and decoupled carbon-water dynamics. Our approach adds to this body of work by leveraging in-footprint sap flow measurements in combination with UAS-derived vegetation maps and biometric scaling parameters to quantify transpiration directly within the EC footprint. This hybrid method provides a species-specific, spatially resolved transpiration estimate and offers an alternative path for ET partitioning in structurally heterogeneous drylands.

2. The use of UAS to scale up sapflow hydraulic measurements for estimating transpiration (T) within an eddy footprint area is novel. Previous efforts with UAS had only produced high-resolution ET mapping (Hoffmann et al., 2016; Simpson et al., 2021), or estimated ET via thermal and visible bands (Claire Brenner and Schulz, 2018). Our approach uniquely leverages UAS technology with GeoAI techniques to enhance the accuracy of T estimations within the eddy footprint.

The results obtained by this method (primarily synthesized in Figure 9) illustrate that T values (including their uncertainty envelopes) appear mostly below the measured eddy-flux ET values across the summer period, which fulfills the condition that $T \leq ET$. Additionally, consistent with previous studies (Jana et al., 2018; Vivoni et al., 2008), evaporation (E) dominates evapotranspiration (ET) (E/ET=56%, T/ET=44%) when precipitation is scarce. After the summer monsoon events, transpiration (T) becomes the dominant contributor (T/ET=63%, E/ET=37%). Overall, across the North American summer monsoon season of 2022, our experiments concluded that both T and E contributed similarly to the whole ET water flux (T/ET=50% and E/ET=50%). This finding agrees with Scott et al. (2015); Good et al. (2015); Ryu et al. (2012); Vivoni et al. (2009) that support that roughly half of the water loss due to evapotranspiration is from plant transpiration, with the other half primarily from soil evaporation, despite the 70% Vs 30% covered areas by bare soil and vegetation respectively.

Several land surface and ecohydrological models, including tRIBS (TIN-based Real-time Integrated Basin Simulator) Ivanov et al., 2004, 2008a,b, partition evapotranspiration (ET) into transpiration (T) and soil evaporation (E) using only vegetation fraction (FVC) and bare soil fraction (FBC) within each spatial unit (e.g., Voronoi element or grid cell). This method assumes that vegetated areas contribute solely to transpiration and bare soil exclusively to evaporation, thereby simplifying ET decomposition based strictly on land cover.

While computationally efficient, this approach neglects important eco-physiological and structural traits—such as stomatal conductance, phenology, and canopy density—that regulate transpiration. Studies including Knipper et al. (2022), Ivanov et al. (2008b), and Ait Hssaine et al. (2020) have also used remote sensing-derived vegetation fractions for ET partitioning, but they often fail to capture functional variability across vegetation types and seasons. As a result, models relying solely on surface cover may overestimate T/ET ratios in sparse or dormant canopies and underestimate soil evaporation under partial or shaded vegetation.

This highlights the need to incorporate not just vegetation structure, but also dynamic vegetation function to improve the realism of ET partitioning, especially in semi-arid and heterogeneous landscapes. In models such as tRIBS, ET is typically partitioned as follows (Equations 7 and 8):

$$T = ET \times FVC \tag{7}$$

$$E = ET \times FBC \quad \text{or} \quad E = ET - T \tag{8}$$

If Equations (7) and (8) are applied to a total ET of 140.2mm with FVC = 0.3 and FBC = 0.7, the resulting estimates are T = 140.2mm×0.3 = 42.06mm and E = 140.2mm×0.7 = 98.14mm. This yields a T/ET ratio of 30% and an E/ET ratio of 70%, which significantly underestimates the role of transpiration compared to our method. The discrepancy arises because this simplified partitioning excludes key variables such as leaf area index (LAI) and other vegetation traits that are explicitly considered in our approach.

Another important point of discussion is the fact that several recent studies examining evapotranspiration (ET) partitioning in semi-arid regions often make a simplifying assumption that the transpiration to evapotranspiration ratio (T/ET) approaches unity (T/ET=1) during the rainy season. While this assumption is convenient for modeling purposes, it overlooks critical ecohydrological dynamics and vegetation constraints inherent to such climates. Our findings challenge this oversimplification, indicating that T/ET rarely exceeds 60–70%, even at peak rainfall, and only under conditions of exceptionally high vegetation cover can values near unity be approached. Moreover, empirical and remote sensing data reinforce that soil evaporation remains a substantial component of ET, even during wetter periods, due to sparse canopy closure and high vapor pressure deficits (Scott et al., 2010; Zhang et al., 2017). Thus, models or studies assuming T/ET = 1 may substantially overstate ecosystem productivity and water-use efficiency, misrepresenting the real hydrological functioning in these sensitive landscapes.

Finally, the results of this research are relevant and foundational in the new era of generative artificial intelligence for several reasons. Firstly, accurate partitioning of ET into its components—transpiration (T) and evaporation (E)—provides high-quality training data for machine learning models, enhancing their predictive accuracy. This detailed data helps us understand the underlying biophysical processes, leading to better model generalization across different environments. Improving model capacity for predicting T, E, and ET for more precise and scalable estimates will be critical for managing water resources, optimizing agricultural practices, and assessing ecosystem health in the face of global warming. By leveraging artificial intelligence, these partitioned observations can be integrated with satellite and UAS data to produce high-resolution, real-time ET estimates, thereby improving decision-making and resource management.

## 6. Conclusions

The following conclusions address the core scientific questions guiding this study, with a focus on transpiration dynamics, species-specific variability, and spatial scaling within a semi-arid eddy covariance footprint:

1. How do daily summer transpiration rates of Mesquite and Creosote individuals of different ages compare? Mesquite shrubs exhibited significantly higher daily transpiration rates (2.84 mm/d) than Creosote bushes (1.78 mm/d) across the North American Monsoon season, yielding a mean transpiration ratio of 1.6:1. This ratio, however, varied with climatic conditions—rising to 1.8:1 during dry spells and falling to 1.3:1 following heavy rainfall. These differences underscore distinct water-use strategies, with Mesquite maintaining more consistent transpiration under drought and Creosote demonstrating stronger physiological response to episodic moisture availability.

2. How can plant-level measurements be integrated for terrain-wide transpiration estimates? The study successfully scaled plant-level transpiration to the ecosystem scale by leveraging biometric scaling laws—relating branch number and size to projected canopy area—and UAS-based classification of vegetation cover. A deep-learning segmentation model (SAM) enabled precise identification and classification of Mesquite and Creosote individuals across the eddy covariance footprint. This integration of ground observations with remote sensing and AI produced a replicable and scalable framework for estimating transpiration over heterogeneous dryland landscapes.

3. What is the contribution of transpiration to total evapotranspiration within the eddy covariance footprint? Across the summer season, plant transpiration contributed approximately 50% of total ET. This proportion was dynamic: during dry periods, the transpiration-to-ET ratio (T/ET) dropped to 44%, reflecting dominant soil evaporation, while following intense monsoon events, it rose to 63% as vegetation accessed moisture from the root zone. These findings challenge common assumptions in ecohydrological models that T/ET approaches unity during wet periods and emphasize the continued importance of soil evaporation, even during high rainfall. The method's performance demonstrates its value for improving ET partitioning accuracy in dryland ecohydrology and for informing land surface and hydrologic models with functionally detailed observations.

## Acknowledgments

The authors gratefully acknowledge Professor Laura Alvarez for providing access to the Phantom 4 unmanned aerial system (UAS) and MicaSense RedEdge camera from the GeoSenSE Lab at UTEP, and for serving as the Pilot in Command during UAS operations. This study was supported and monitored by The National Oceanic and Atmospheric Administration – Cooperative Science Center for Earth System Sciences and Remote Sensing Technologies under the Cooperative Agreement Grant #: NA22SEC4810016. The authors would like to thank the NOAA Office of Education, the Educational Partnership Program with Minority Serving Institutions (NOAA-EPP/MSI) and the NOAA-CESSRST-II for full fellowship support for Stephanie Marquez. The study was also supported by the Army Research Office (ARO) Cooperative Agreement Number W911NF-24-1-0296 that supported Dr. Hernan Moreno and Dr. Laura Alvarez. The statements, findings, conclusions, and recommendations are those of the author(s) and do not necessarily reflect the views of NOAA or the ARO.

## References

Ait Hssaine, B., Merlin, O., Ezzahar, J., Ojha, N., Er-Raki, S., Khabba, S., 2020. An evapotranspiration model self-calibrated from remotely sensed surface soil moisture, land surface temperature and vegetation cover fraction: application to disaggregated smos

and modis data. Hydrology and Earth System Sciences 24, 1781–1803. doi:`10.5194/hess-24-1781-2020`.

Alam, M., Lamb, D., Warwick, N., 2021. A canopy transpiration model based on scaling up stomatal conductance and radiation interception as affected by leaf area index. Water 13, 252. doi:`10.3390/w13030252`.

Ansley, R.J., Price, D.L., Dowhower, S.L., Carlson, D.H., 1992. Seasonal trends in leaf area of honey mesquite trees: Determination using image analysis. Journal of Range Management 45, 339–344. URL: `https://doi.org/10.2307/4003079`, doi:`10.2307/4003079`.

Baker, J.M., Van Bavel, C.H.M., 1987. Measurement of mass flow of water in the stems of herbaceous plants. Plant, Cell & Environment 10, 777–782. URL: `https://onlinelibrary.wiley.com/doi/abs/10.1111/1365-3040.ep11604765`, doi:`https://doi.org/10.1111/1365-3040.ep11604765`.

Bergametti, G., Gillette, D.A., 2010. Aeolian sediment fluxes measured over various plant/soil complexes in the chihuahuan desert. Journal of Geophysical Research: Earth Surface 115, F03044. URL: `https://agupubs.onlinelibrary.wiley.com/doi/abs/10.1029/2009JF001543`, doi:`https://doi.org/10.1029/2009JF001543`.

Berry, Z., Looker, N., Holwerda, F., Ortiz-Colín, P., González Martínez, T., Asbjornsen, H., 2017. Why size matters: The interactive influences of tree diameter distribution and sap flow parameters on upscaled transpiration. Tree Physiology 38, 263–275. doi:`10.1093/treephys/tpx124`.

Browning, D.M., Karl, J.W., Morin, D., Richardson, A.D., Tweedie, C.E., 2017. Phenocams bridge the gap between field and satellite observations in an arid grassland ecosystem. Remote Sensing 9, 1071. URL: `https://www.mdpi.com/2072-4292/9/10/1071`, doi:`10.3390/rs9101071`.

Bu, J., Gan, G., Chen, J., Su, Y., Yuan, M., Gao, Y., Domingo, F., López-Ballesteros, A., Migliavacca, M., El-Madany, T., Gentine, P., Xiao, J., Garcia, M., 2024. Dryland evapotranspiration from remote sensing solar-induced chlorophyll fluorescence: Constraining an optimal stomatal model within a two-source energy balance model. Remote Sensing of Environment 303, 113999. doi:10.1016/j.rse.2024.113999.

Burgess, S., Dawson, T., 2007. Using branch and basal trunk sap flow measurements to estimate whole-plant water capacitance: A caution. Plant and Soil 305, 5–13. doi:10.1007/s11104-007-9378-2.

Carlson, D.H., Thurow, T.L., Knight, R.W., Heitschmidt, R.K., 1990. Effect of Honey Mesquite on the Water Balance of Texas Rolling Plains Rangeland. Journal of Range Management 43, 491–496. URL: https://doi.org/10.2307/4002351, doi:10.2307/4002351.

Cavanaugh, M.L., Kurc, S.A., Scott, R.L., 2011. Evapotranspiration partitioning in semiarid shrubland ecosystems: a two-site evaluation of soil moisture control on transpiration. Ecohydrology 4, 671–681. URL: https://onlinelibrary.wiley.com/doi/abs/10.1002/eco.157, doi:https://doi.org/10.1002/eco.157.

Chopping, M.J., Rango, A., Havstad, K.M., Schiebe, F.R., Ritchie, J.C., Schmugge, T.J., French, A.N., Su, L., McKee, L., Davis, M., 2003. Canopy attributes of desert grassland and transition communities derived from multiangular airborne imagery. Remote Sensing of Environment 85, 339–354. URL: https://www.sciencedirect.com/science/article/pii/S0034425703000129, doi:https://doi.org/10.1016/S0034-4257(03)00012-9.

Claire Brenner, Matthias Zeeman, M.B., Schulz, K., 2018. Estimation of evapotranspiration of temperate grassland based on high-resolution thermal and visible range imagery from unmanned aerial systems. International Journal of Remote Sensing 39, 5141–5174. doi:10.1080/01431161.2018.1471550.

Darrouzet-Nardi, A., Asaff, I.S., Mauritz, M., Roman, K., Keats, E., Tweedie, C.E.,

McLaren, J.R., 2023. Consistent microbial and nutrient resource island patterns during monsoon rain in a chihuahuan desert bajada shrubland. Ecosphere 14, e4475. URL: https://esajournals.onlinelibrary.wiley.com/doi/abs/10.1002/ecs2.4475, doi:https://doi.org/10.1002/ecs2.4475.

Dawson, T.E., Ehleringer, J.R., 1991. Streamside trees that do not use stream water. Nature 350, 335–337. URL: https://doi.org/10.1038/350335a0, doi:10.1038/350335a0.

Denager, T., Looms, M.C., Sonnenborg, T.O., Jensen, K.H., 2020. Comparison of evapotranspiration estimates using the water balance and the eddy covariance methods. Vadose Zone Journal 19, e20032. URL: https://acsess.onlinelibrary.wiley.com/doi/abs/10.1002/vzj2.20032, doi:https://doi.org/10.1002/vzj2.20032.

Deng, Q., Hui, D., Chu, G., Han, X., Zhang, Q., 2017. Rain-induced changes in soil co2 flux and microbial community composition in a tropical forest of china. Scientific Reports 7, 5539. doi:10.1038/s41598-017-06345-2.

Donovan, L., Ehleringer, J., 1991. Ecophysiological differences among juvenile and reproductive plants of several woody species. Oecologia 86, 594–597. doi:10.1007/BF00318327.

Dugas, W.A., Hicks, R.A., Wright, P., 1998. Effect of removal ofJuniperus asheion evapotranspiration and runoff in the Seco Creek Watershed. Water Resources Research 34, 1499–1506. URL: https://agupubs.onlinelibrary.wiley.com/doi/abs/10.1029/98WR00556, doi:https://doi.org/10.1029/98WR00556.

Dukat, P., Ziemblińska, K., Räsänen, M., Olejnik, J., Urbaniak, M., 2023. Scots pine responses to drought investigated with eddy covariance and sap flow methods. European Journal of Forest Research. 142, 671–690. doi:10.1007/s10342-023-01549-w.

Duniway, M., Herrick, J., Monger, C., 2009. Spatial and temporal variability of plant-available water in calcium carbonate-cemented soils and consequences for arid ecosystem resilience. Oecologia 163, 215–26. doi:10.1007/s00442-009-1530-7.

Eamus, D., Boulain, N., Cleverly, J., Breshears, D.D., 2013. Global change-type drought-induced tree mortality: vapor pressure deficit is more important than temperature per se in causing decline in tree health. Ecology and Evolution 3, 2711–2729. URL: https://onlinelibrary.wiley.com/doi/abs/10.1002/ece3.664, doi:https://doi.org/10.1002/ece3.664.

Ehleringer, J., Phillips, S., Schuster, W., Sandquist, D., 1991. Differential utilization of summer rains by desert plants. Oecologia 88, 430–434. doi:10.1007/BF00317589.

Ehleringer, J.R., Roden, J., Dawson, T.E., 2000. Assessing Ecosystem-Level Water Relations Through Stable Isotope Ratio Analyses. Springer New York, New York, NY. pp. 181–198. URL: https://doi.org/10.1007/978-1-4612-1224-9_13, doi:10.1007/978-1-4612-1224-9_13.

Eichelmann, E., Mantoani, M., Chamberlain, S., Hemes, K., Oikawa, P., Szutu, D., Valach, A., Verfaillie, J., Baldocchi, D., 2021. A novel approach to partitioning evapotranspiration into evaporation and transpiration in flooded ecosystems. Global Change Biology 28, 990–1007. doi:10.1101/2021.04.06.438244.

Ewers, B., Mackay, D., Tang, J., Bolstad, P., Samanta, S., 2008. Intercomparison of sugar maple (acer saccharum marsh.) stand transpiration responses to environmental conditions from the western great lakes region of the united states. Agricultural and Forest Meteorology 148, 231–246. URL: https://www.sciencedirect.com/science/article/pii/S0168192307002079, doi:https://doi.org/10.1016/j.agrformet.2007.08.003. chequamegon Ecosystem-Atmosphere Study Special Issue: Ecosystem-Atmosphere Carbon and Water Cycling in the Temperate Northern Forests of the Great Lakes Region.

Ferretti, D., Pendall, E., Morgan, J., Nelson, J., Lecain, D., Mosier, A., 2003. Partitioning evapotranspiration fluxes from a colorado grassland using stable isotopes: Seasonal varia-tions and ecosystem implications of elevated atmospheric co2. Plant and Soil 254, 291–303. doi:10.1023/A:1025511618571.

Flores, A., 2021. Linking Plot and Landscape Level Phenology. Master's thesis. University of Texas at El Paso. URL: https://scholarworks.utep.edu/open_etd/3408.

French, A., Schmugge, T., Ritchie, J., Hsu, A., Jacob, F., Ogawa, K., 2008. Detecting land cover change at the jornada experimental range, new mexico with aster emissivities. Remote Sensing of Environment 112, 1730–1748. URL: https://www.sciencedirect.com/science/article/pii/S003442570700418X, doi:https://doi.org/10.1016/j.rse.2007.08.020.

Gabriel, C.E., Kellman, L., 2014. Investigating the role of moisture as an environmental constraint in the decomposition of shallow and deep mineral soil organic matter of a temperate coniferous soil. Soil Biology and Biochemistry 68, 373–384. URL: https://www.sciencedirect.com/science/article/pii/S0038071713003441, doi:https://doi.org/10.1016/j.soilbio.2013.10.009.

Gibbens, R.P., Hicks, R.A., Dugasf, W.A., 1996. Structure and function of c and c, chihuahuan desert plant communities. standing crop and leaf. Journal of Arid Environments 34, 47–62. URL: https://www.sciencedirect.com/science/article/pii/S0140196396900920, doi:https://doi.org/10.1006/jare.1996.0092.

Good, S.P., Noone, D., Kurita, N., Benetti, M., Bowen, G.J., 2015. D/h isotope ratios in the global hydrologic cycle. Geophysical Research Letters 42, 5042–5050. URL: https://agupubs.onlinelibrary.wiley.com/doi/abs/10.1002/2015GL064117, doi:https://doi.org/10.1002/2015GL064117.

Gough, C.M., Vogel, C.S., Schmid, H.P., Curtis, P.S., 2008. Controls on Annual Forest Carbon Storage: Lessons from the Past and Predictions for the Future. BioScience 58, 609–622. URL: https://doi.org/10.1641/B580708, doi:10.1641/B580708.

Goulden, M.L., Daube, B.C., Fan, S.M., Sutton, D.J., Bazzaz, A., Munger, J.W., Wofsy, S.C., 1997. Physiological responses of a black spruce forest to weather. Journal of Geophysical Research: Atmospheres 102, 28987–28996. URL: https://agupubs.onlinelibrary.wiley.com/doi/abs/10.1029/97JD01111, doi:https://doi.org/10.1029/97JD01111.

Granier, A., Biron, P., Lemoine, D., 2000. Water balance, transpiration and canopy conductance in two beech stands. Agricultural and Forest Meteorology 100, 291–308. URL: https://www.sciencedirect.com/science/article/pii/S0168192399001513, doi:https://doi.org/10.1016/S0168-1923(99)00151-3.

Hamerlynck, E., Scott, R., Moran, M., Schwander, A., Connor, E., Huxman, T., 2011. Inter- and under-canopy soil water, leaf-level and whole-plant gas exchange dynamics of a semi-arid perennial c4 grass. Oecologia 165, 17–29. doi:10.1007/s00442-010-1757-3.

Havstad, K., Kustas, W., Rango, A., Ritchie, J., Schmugge, T., 2000. Jornada experimental range: A unique arid land location for experiments to validate satellite systems. Remote Sensing of Environment 74, 13–25. URL: https://www.sciencedirect.com/science/article/pii/S0034425700001188, doi:https://doi.org/10.1016/S0034-4257(00)00118-8.

Havstad, K.M., Huenneke, L.F., Schlesinger, W.H., 2006. Structure and Function of a Chihuahuan Desert Ecosystem: The Jornada Basin Long-Term Ecological Research Site. Oxford University Press. URL: https://doi.org/10.1093/oso/9780195117769.001.0001, doi:10.1093/oso/9780195117769.001.0001.

Herbst, M., Rosier, P., McNeil, D., Harding, R., 2008. Seasonal variability of interception evaporation from the canopy of a mixed deciduous forest. Agricultural and Forest Meteorology 148, 1655–1667. doi:10.1016/j.agrformet.2008.05.011.

Hoffmann, H., Jensen, R., Thomsen, A., Nieto, H., Rasmussen, J., Friborg, T., 2016. Crop

water stress maps for an entire growing season from visible and thermal uav imagery. Biogeosciences 13, 6545–6563. doi:10.5194/bg-13-6545-2016.

Hultine, K., Scott, R., Cable, W., Goodrich, D., Williams, D., 2004. Hydraulic redistribution by a dominant, warm-desert phreatophyte: Seasonal patterns and response to precipitation pulses. Functional Ecology 18, 530 – 538. doi:10.1111/j.0269-8463.2004.00867.x.

Humphreys, E., Black, T., Ethier, G., Drewitt, G., Spittlehouse, D., Jork, E.M., Nesic, Z., Livingston, N., 2003. Annual and seasonal variability of sensible and latent heat fluxes above a coastal douglas-fir forest, british columbia, canada. Agricultural and Forest Meteorology 115, 109–125. URL: https://www.sciencedirect.com/science/article/pii/S0168192302001715, doi:https://doi.org/10.1016/S0168-1923(02)00171-5. a tribute to George W. Thurtell's contributions in micrometeorology.

Huxman, T., Snyder, K., Tissue, D., Leffler, A., Ogle, K., Pockman, W., Sandquist, D., Potts, D., Schwinning, S., 2004. Precipitation pulses and carbon fluxes in semiarid and arid ecosystems. Oecologia 141, 254–268. doi:10.1007/s00442-004-1682-4.

Huxman, T.E., Wilcox, B.P., Breshears, D.D., Scott, R.L., Snyder, K.A., Small, E.E., Hultine, K., Pockman, W.T., Jackson, R.B., 2005. Ecohydrological implications of woody plant encroachment. Ecology 86, 308–319. URL: https://esajournals.onlinelibrary.wiley.com/doi/abs/10.1890/03-0583, doi:https://doi.org/10.1890/03-0583.

Ivanov, V.Y., Bras, R.L., Vivoni, E.R., 2008a. Vegetation-hydrology dynamics in complex terrain of semiarid areas: 1. a mechanistic approach to modeling dynamic feedbacks. Water Resources Research 44, W03429. URL: https://agupubs.onlinelibrary.wiley.com/doi/abs/10.1029/2006WR005588, doi:https://doi.org/10.1029/2006WR005588.

Ivanov, V.Y., Bras, R.L., Vivoni, E.R., 2008b. Vegetation-hydrology dynamics in complex terrain of semiarid areas: 2. energy-water controls of vegetation spatiotemporal dynamics and topographic niches of favorability. Water Resources Research 44, W03430. URL: https:

//agupubs.onlinelibrary.wiley.com/doi/abs/10.1029/2006WR005595, doi:https://doi.org/10.1029/2006WR005595.

Ivanov, V.Y., Vivoni, E.R., Bras, R.L., Entekhabi, D., 2004. Preserving high-resolution surface and rainfall data in operational-scale basin hydrology: a fully-distributed physically-based approach. Journal of Hydrology 298, 80–111. URL: https://www.sciencedirect.com/science/article/pii/S0022169404002409, doi:https://doi.org/10.1016/j.jhydrol.2004.03.041. the Distributed Model Intercomparison Project (DMIP).

Jana, S., Rajagopalan, B., Alexander, M.A., Ray, A.J., 2018. Understanding the dominant sources and tracks of moisture for summer rainfall in the southwest united states. Journal of Geophysical Research: Atmospheres 123, 4850–4870. URL: https://agupubs.onlinelibrary.wiley.com/doi/abs/10.1029/2017JD027652, doi:https://doi.org/10.1029/2017JD027652.

Jasechko, S., Sharp, Z., Gibson, J., Birks, S., Yi, Y., Fawcett, P., 2013. Terrestrial water fluxes dominated by transpiration. Nature 496, 347–350. doi:10.1038/nature11983.

Javadian, M., Aubrecht, D.M., Fisher, J.B., Scott, R.L., Burns, S.P., Diehl, J.L., Munger, J.W., Richardson, A.D., 2024. Scaling individual tree transpiration with thermal cameras reveals interspecies differences to drought vulnerability. Geophysical Research Letters 51, e2024GL111479. URL: https://agupubs.onlinelibrary.wiley.com/doi/abs/10.1029/2024GL111479, doi:https://doi.org/10.1029/2024GL111479.

Jensen, K.H., Berg-Sørensen, K., Bruus, H., Holbrook, N.M., Liesche, J., Schulz, A., Zwieniecki, M.A., Bohr, T., 2016. Sap flow and sugar transport in plants. Rev. Mod. Phys. 88, 035007. URL: https://link.aps.org/doi/10.1103/RevModPhys.88.035007, doi:10.1103/RevModPhys.88.035007.

Jha, C., Thumaty, K.C., Reddy, S., Sonakia, A., Dadhwal, V., 2013. Analysis of carbon

dioxide, water vapour and energy over an indian teak mixed deciduous forest for winter and summer months using eddy covariance technique. Journal of Earth System Science 122, 1259–1268. doi:10.1007/s12040-013-0350-7.

Katul, G., Ellsworth, D., Lai, C.T., 2000. Modeling assimilation and intercellular co2 from measured conductance. Plant, Cell & Environment 23, 1313–1328. doi:10.1046/j.1365-3040.2000.00641.x.

Kim, S., Kiniry, J., Loomis, L., 2017. Creosote bush, an arid zone survivor in southwestern u.s.: 1. identification of morphological and environmental factors that affect its growth and development. Journal of Agriculture and Ecology Research International 11, 1–14. doi:10.9734/JAERI/2017/33204.

Kirillov, A., Mintun, E., Ravi, N., Mao, H., Rolland, C., Gustafson, L., Xiao, T., Whitehead, S., Berg, A.C., Lo, W.Y., Dollár, P., Girshick, R., 2023. Segment anything. URL: https://arxiv.org/abs/2304.02643.

Kjelgaard, J., Stockle, C., Black, R., Campbell, G., 1997. Measuring sap flow with the heat balance approach using constant and variable heat inputs. Agricultural and Forest Meteorology 85, 239–250. URL: https://www.sciencedirect.com/science/article/pii/S0168192396023970, doi:https://doi.org/10.1016/S0168-1923(96)02397-0.

Kljun, N., Calanca, P., Rotach, M.W., Schmid, H.P., 2004. A simple parameterisation for flux footprint predictions. Boundary-Layer Meteorology 112, 503–523. URL: https://api.semanticscholar.org/CorpusID:122184822, doi:10.1023/B:BOUN.0000030653.71031.96.

Knipper, K., Anderson, M., Bambach, N., Kustas, W., Zahn, E., Hain, C., McElrone, A., Belfiore, O.R., Castro, S., Alsina, M., Saa, S., 2022. Evaluation of partitioned evaporation and transpiration estimates within the disalexi modeling framework over irrigated crops in california. Remote Sensing 15, 68. doi:10.3390/rs15010068.

Kraimer, R.A., Monger, H.C., Steiner, R.L., 2005. Mineralogical distinctions of carbonates in desert soils. Soil Science Society of America Journal 69, 1773–1781. URL: https://acsess.onlinelibrary.wiley.com/doi/abs/10.2136/sssaj2004.0275, doi:https://doi.org/10.2136/sssaj2004.0275.

Kurc, S.A., Small, E.E., 2004. Dynamics of evapotranspiration in semiarid grassland and shrubland ecosystems during the summer monsoon season, central New Mexico. Water Resources Research 40, W09305. URL: https://agupubs.onlinelibrary.wiley.com/doi/abs/10.1029/2004WR003068, doi:https://doi.org/10.1029/2004WR003068.

Kurc, S.A., Small, E.E., 2007. Soil moisture variations and ecosystem-scale fluxes of water and carbon in semiarid grassland and shrubland. Water Resources Research 43, W06416. URL: https://agupubs.onlinelibrary.wiley.com/doi/abs/10.1029/2006WR005011, doi:https://doi.org/10.1029/2006WR005011.

Lascano, R.J., Baumhardt, R.L., Lipe, W.N., 1992. Measurement of water flow in young grapevines using the stem heat balance method. American Journal of Enology and Viticulture 43, 159–165. URL: https://www.ajevonline.org/content/43/2/159, doi:10.5344/ajev.1992.43.2.159.

Law, B.E., Waring, R.H., 1994. Combining remote sensing and climatic data to estimate net primary production across oregon. Ecological Applications 4, 717–728. URL: https://esajournals.onlinelibrary.wiley.com/doi/abs/10.2307/1942002, doi:https://doi.org/10.2307/1942002.

Li, X., Gentine, P., Lin, C., Zhou, S., Sun, Z., Zheng, Y., Liu, J., Zheng, C., 2019. A simple and objective method to partition evapotranspiration into transpiration and evaporation at eddy-covariance sites. Agricultural and Forest Meteorology 265, 171–182. doi:10.1016/j.agrformet.2018.11.017.

Loik, M.E., Breshears, D.D., Lauenroth, W.K., Belnap, J., 2004. A multi-scale perspective of water pulses in dryland ecosystems: climatology and ecohydrology of the western USA. Oecologia 141, 269–281. doi:10.1007/s00442-004-1570-y.

Marion, W.H.S., Fonteyn, P.J., 1990. Spatial variability of caco3 solubility in a chihuahuan desert soil. Arid Soil Research and Rehabilitation 4, 181–191. doi:10.1080/15324989009381247.

Marquez, S.N., 2023. Quantifying The Contribution Of Atmospheric And Land Surface Characteristics To The Prediction Of Sub-Pixel Scale Surface Soil Moisture In The Jornada Experimental Range Through Interpretable Machine Learning. Master's thesis. University of Texas at El Paso. URL: https://scholarworks.utep.edu/open_etd/3923.

Mayo, J., 2023. Using Machine Learning And Distributed Hydrologic Modeling To Predict Soil Texture, Surface Soil Moisture And Evapotranspiration In Jornada Experimental Range, Southwestern U.s. Master's thesis. University of Texas at El Paso. URL: https://scholarworks.utep.edu/open_etd/3924.

McClaran, M., Van Devender, T., 1997. The Desert Grassland. University of Arizona Press. URL: https://books.google.com/books?id=BAlCcnXtXg4C.

Monteith, J., 1965. Evaporation and environment. Symposia of the Society for Experimental Biology 19, 205—234. URL: https://pubmed.ncbi.nlm.nih.gov/5321565/.

Moran, M., Scott, R., Keefer, T., Emmerich, W., Hernandez, M., Nearing, G., Paige, G., Cosh, M., O'Neill, P., 2009. Partitioning evapotranspiration in semiarid grassland and shrubland ecosystems using time series of soil surface temperature. Agricultural and Forest Meteorology 149, 59–72. URL: https://www.sciencedirect.com/science/article/pii/S0168192308002013, doi:https://doi.org/10.1016/j.agrformet.2008.07.004.

Nelson, J.A., Pérez-Priego, O., Zhou, S., Poyatos, R., Zhang, Y., Blanken, P.D., Gimeno, T.E., Wohlfahrt, G., Desai, A.R., Gioli, B., Limousin, J.M., Bonal, D., Paul-Limoges, E.,

Scott, R.L., Varlagin, A., Fuchs, K., Montagnani, L., Wolf, S., Delpierre, N., Berveiller, D., Gharun, M., Belelli Marchesini, L., Gianelle, D., Šigut, L., Mammarella, I., Siebicke, L., Andrew Black, T., Knohl, A., Hörtnagl, L., Magliulo, V., Besnard, S., Weber, U., Carvalhais, N., Migliavacca, M., Reichstein, M., Jung, M., 2020. Ecosystem transpiration and evaporation: Insights from three water flux partitioning methods across fluxnet sites. Global Change Biology 26, 6916–6930. URL: https://onlinelibrary.wiley.com/doi/abs/10.1111/gcb.15314, doi:https://doi.org/10.1111/gcb.15314.

Niklas, K., 1994. Plant Allometry: The Scaling of Form and Process. Women in Culture and Society, University of Chicago Press. URL: https://books.google.com/books?id=2th19CVNWtcC.

Nobles, M., Wilding, L., Lin, H., 2010. Flow pathways of bromide and brilliant blue fcf tracers in caliche soils. Journal of Hydrology 393, 114–122. doi:10.1016/j.jhydrol.2010.03.014.

Novick, K., Ficklin, D., Stoy, P., Williams, C., Bohrer, G., Oishi, A., Papuga, S., Blanken, P., Noormets, A., Sulman, B., Scott, R., Wang, L., Phillips, R., 2016. The increasing importance of atmospheric demand for ecosystem water and carbon fluxes. Nature Climate Change 6, 1023–1027. doi:10.1038/nclimate3114.

Novick, K., Oren, R., Stoy, P., Siqueira, M., Katul, G., 2009. Nocturnal evapotranspiration in eddy-covariance records from three co-located ecosystems in the southeastern u.s.: Implications for annual fluxes. Agricultural and Forest Meteorology 149, 1491–1504. URL: https://www.sciencedirect.com/science/article/pii/S0168192309000914, doi:https://doi.org/10.1016/j.agrformet.2009.04.005.

Ogle, K., Reynolds, J., 2004. Plant responses to precipitation in desert ecosystems: Integrating functional types, pulses, thresholds, and delays. Oecologia 141, 282–94. doi:10.1007/s00442-004-1507-5.

Oishi, A., Oren, R., Stoy, P., 2008. Estimating components of forest evapotranspiration: A footprint approach for scaling sap fux measurements. Agricultural and Forest Meteorology 148, 1719–1732. doi:10.1016/j.agrformet.2008.06.013.

Oren, R., Ellsworth, D.S., Johnsen, K.H., Phillips, N., Ewers, B.E., Maier, C.A., Schäfer, K.V.R., McCarthy, H., Hendrey, G., McNulty, S.G., Katul, G.G., 2001. Soil fertility limits carbon sequestration by forest ecosystems in a co2-enriched atmosphere. Nature 411, 469–472. doi:10.1038/35078064.

Oren, R., Sperry, J.S., Katul, G.G., Pataki, D.E., Ewers, B.E., Phillips, N., Schäfer, K.V.R., 1999. Survey and synthesis of intra- and interspecific variation in stomatal sensitivity to vapour pressure deficit. Plant, Cell & Environment 22, 1515–1526. URL: https://onlinelibrary.wiley.com/doi/abs/10.1046/j.1365-3040.1999.00513.x, doi:https://doi.org/10.1046/j.1365-3040.1999.00513.x.

Petrie, M., Collins, S., Gutzler, D., Moore, D., 2014. Regional trends and local variability in monsoon precipitation in the northern Chihuahuan Desert, USA. Journal of Arid Environments 103, 63–70. URL: https://www.sciencedirect.com/science/article/pii/S014019631400010X, doi:https://doi.org/10.1016/j.jaridenv.2014.01.005.

Potts, D.L., Huxman, T.E., Cable, J.M., English, N.B., Ignace, D.D., Eilts, J.A., Mason, M.J., Weltzin, J.F., Williams, D.G., 2006. Antecedent moisture and seasonal precipitation influence the response of canopyscale carbon and water exchange to rainfall pulses in a semiarid grassland. New Phytologist 170, 849–860. URL: https://nph.onlinelibrary.wiley.com/doi/abs/10.1111/j.1469-8137.2006.01732.x, doi:https://doi.org/10.1111/j.1469-8137.2006.01732.x.

Pu, J., Yan, K., Zhou, G., Lei, Y., Zhu, Y., Guo, D., Li, H., Xu, L., Knyazikhin, Y., Myneni, R.B., 2020. Evaluation of the modis lai/fpar algorithm based on 3d-rtm simulations: A case study of grassland. Remote Sensing 12, 3391. URL: https://www.mdpi.com/2072-4292/12/20/3391, doi:10.3390/rs12203391.

Pérez-Priego, O., Katul, G., Reichstein, M., El-Madany, T., Ahrens, B., Carrara, A., Scanlon, T., Migliavacca, M., 2018. Partitioning eddy covariance water flux components using physiological and micrometeorological approaches. Journal of Geophysical Research: Biogeosciences 123, 3353–3370. doi:10.1029/2018JG004637.

Reich, E., Samuels-Crow, K., Bradford, J., Litvak, M., Schlaepfer, D., Ogle, K., 2024. A semi-mechanistic model for partitioning evapotranspiration reveals transpiration dominates the water flux in drylands. Journal of Geophysical Research: Biogeosciences 129, e2023JG007914. doi:10.1029/2023JG007914.

Reynolds, J., Kemp, P., Ogle, K., Fernández, R., 2004. Modifying the 'pulse-reserve' paradigm for deserts of north america: Precipitation pulses, soil water, and plant responses. Oecologia 141, 194–210. doi:10.1007/s00442-004-1524-4.

Romig, D., Munk, L., Stein, T., 2006. Leaf area and root density measurements for use in cover performance evaluations on semi-arid reclaimed mine lands. Journal American Society of Mining and Reclamation 2006, 1694–1703. doi:10.21000/JASMR06021694.

Ryu, Y., Baldocchi, D., Black, T., Detto, M., Law, B., Leuning, R., Miyata, A., Reichstein, M., Vargas, R., Ammann, C., Beringer, J., Flanagan, L., Gu, L., Hutley, L., Kim, J., Mccaughey, H., Moors, E., Rambal, S., Vesala, T., 2012. On the temporal upscaling of evapotranspiration from instantaneous remote sensing measurements to 8-day mean daily-sums. Agricultural and Forest Meteorology 152, 212–222. doi:10.1016/j.agrformet.2011.09.010.

Sakuratani, T., 1981. A heat balance method for measuring water flux in the stem of intact plants. Journal of Agricultural Meteorology 37, 9–17. doi:10.2480/agrmet.37.9.

Scanlon, T., Schmidt, D., Skaggs, T., 2019. Correlation-based flux partitioning of water vapor and carbon dioxide fluxes: Method simplification and estimation of canopy water use efficiency. Agricultural and Forest Meteorology 279, 107732. doi:10.1016/j.agrformet.2019.107732.

Schenk, H.J., Jackson, R.B., 2002. Rooting depths, lateral root spreads and below-ground/above-ground allometries of plants in water-limited ecosystems. Journal of Ecology 90, 480–494. URL: https://besjournals.onlinelibrary.wiley.com/doi/abs/10.1046/j.1365-2745.2002.00682.x, doi:https://doi.org/10.1046/j.1365-2745.2002.00682.x.

Schlesinger, W.H., Jasechko, S., 2014. Transpiration in the global water cycle. Agricultural and Forest Meteorology 189-190, 115–117. URL: https://www.sciencedirect.com/science/article/pii/S0168192314000203, doi:https://doi.org/10.1016/j.agrformet.2014.01.011.

Schwinning, S., Sala, O., Loik, M., Ehleringer, J., 2004. Thresholds, memory, and seasonality: Understanding pulse dynamics in arid/semi-arid ecosystems. Oecologia 141, 191–193. doi:10.1007/s00442-004-1683-3.

Schwinning, S., Starr, B.I., Ehleringer, J.R., 2003. Dominant cold desert plants do not partition warm season precipitation by event size. Oecologia 136, 252–260. doi:10.1007/s00442-003-1255-y.

Scott, R., Biederman, J., 2017. Partitioning evapotranspiration using long-term carbon dioxide and water vapor fluxes: New approach to et partitioning. Geophysical Research Letters 44, 6833–6840. doi:10.1002/2017GL074324.

Scott, R., Biederman, J., Hamerlynck, E., Barron-Gafford, G., 2015. The carbon balance pivot point of southwestern u.s. semiarid ecosystems: Insights from the 21st century drought. Journal of Geophysical Research: Biogeosciences 120, 2612–2624. doi:10.1002/2015JG003181.

Scott, R.L., 2010. Using watershed water balance to evaluate the accuracy of eddy covariance evaporation measurements for three semiarid ecosystems. Agricultural and Forest

Meteorology 150, 219–225. URL: https://www.sciencedirect.com/science/article/pii/S0168192309002512, doi:https://doi.org/10.1016/j.agrformet.2009.11.002.

Scott, R.L., Hamerlynck, E.P., Jenerette, G.D., Moran, M.S., Barron-Gafford, G.A., 2010. Carbon dioxide exchange in a semidesert grassland through drought-induced vegetation change. Journal of Geophysical Research: Biogeosciences 115, G03026. URL: https://agupubs.onlinelibrary.wiley.com/doi/abs/10.1029/2010JG001348, doi:https://doi.org/10.1029/2010JG001348.

Scott, R.L., Huxman, T.E., Cable, W.L., Emmerich, W.E., 2006a. Partitioning of evapotranspiration and its relation to carbon dioxide exchange in a Chihuahuan Desert shrubland. Hydrological Processes 20, 3227–3243. URL: https://onlinelibrary.wiley.com/doi/abs/10.1002/hyp.6329, doi:https://doi.org/10.1002/hyp.6329.

Scott, R.L., Huxman, T.E., Williams, D.G., Goodrich, D.C., 2006b. Ecohydrological impacts of woody-plant encroachment: seasonal patterns of water and carbon dioxide exchange within a semiarid riparian environment. Global Change Biology 12, 311–324. URL: https://onlinelibrary.wiley.com/doi/abs/10.1111/j.1365-2486.2005.01093.x, doi:https://doi.org/10.1111/j.1365-2486.2005.01093.x.

Scott, R.L., Knowles, J.F., Nelson, J.A., Gentine, P., Li, X., Barron-Gafford, G., Bryant, R., Biederman, J.A., 2021. Water availability impacts on evapotranspiration partitioning. Agricultural and Forest Meteorology 297, 108251. URL: https://www.sciencedirect.com/science/article/pii/S0168192320303531, doi:https://doi.org/10.1016/j.agrformet.2020.108251.

Senock, R.S., Ham, J.M., 1993. Heat balance sap flow gauge for small diameter stems. Plant, Cell & Environment 16, 593–601. URL: https://onlinelibrary.wiley.com/doi/abs/10.1111/j.1365-3040.1993.tb00908.x, doi:https://doi.org/10.1111/j.1365-3040.1993.tb00908.x.

Serna Pérez, A., Monger, C., Herrick, J., Murray, L., 2006. Carbon dioxide emissions from exhumed petrocalcic horizons. Soil Science Society of America Journal 70, 795–805. doi:10.2136/sssaj2005.0099.

Simpson, J.E., Holman, F., Nieto, H., Voelksch, I., Mauder, M., Klatt, J., Fiener, P., Kaplan, J.O., 2021. High spatial and temporal resolution energy flux mapping of different land covers using an off-the-shelf unmanned aerial system. Remote Sensing 13, 1286. URL: https://www.mdpi.com/2072-4292/13/7/1286, doi:10.3390/rs13071286.

Skaggs, T., Anderson, R., Alfieri, J., Scanlon, T., Kustas, W., 2018. Fluxpart: Open source software for partitioning carbon dioxide and water vapor fluxes. Agricultural and Forest Meteorology 253, 218–224. doi:10.1016/j.agrformet.2018.02.019.

Smith, D., Allen, S., 1996. Measurement of sap flow in plant stems. Journal of Experimental Botany 47, 1833–1844. URL: https://doi.org/10.1093/jxb/47.12.1833, doi:10.1093/jxb/47.12.1833.

Stoy, P., El-Madany, T., Fisher, J., Gentine, P., Gerken, T., Good, S., Klosterhalfen, A., Liu, S., Miralles, D., Pérez-Priego, O., Rigden, A., Skaggs, T., Wohlfahrt, G., Anderson, R., Coenders-Gerrits, M., Jung, M., Maes, W., Mammarella, I., Mauder, M., Wolf, S., 2019. Reviews and syntheses: Turning the challenges of partitioning ecosystem evaporation and transpiration into opportunities. Biogeosciences 16, 3747–3775. doi:10.5194/bg-16-3747-2019.

Sun, X., Wilcox, B.P., Zou, C.B., 2019. Evapotranspiration partitioning in dryland ecosystems: A global meta-analysis of in situ studies. Journal of Hydrology 576, 123–136. doi:10.1016/j.jhydrol.2019.06.022.

Sutanto, S., Hurk, B., Dirmeyer, P., Seneviratne, S., Röckmann, T., Trenberth, K., Blyth, E., Wenninger, J., Hoffmann, G., 2014. Hess opinions "a perspective on isotope versus non-isotope approaches to determine the contribution of transpiration to total evaporation". Hydrology and Earth System Sciences 18, 2815–2827. doi:10.5194/hess-18-2815-2014.

Szutu, D., Papuga, S., 2019. Year-round transpiration dynamics linked with deep soil moisture in a warm desert shrubland. Water Resources Research 55, 5679–5695. doi:10.1029/2018WR023990.

Tang, J., Bolstad, P., Desai, A., Martin, J., Cook, B., Davis, K., 2004. Ecosystem respiration and its components in an old-growth and a mature northern forest. Agricultural and Forest Meteorology , 171–185URL: https://www.sciencedirect.com/science/article/pii/S0168192307002122, doi:https://doi.org/10.1016/j.agrformet.2007.08.008.

Terminel, M., Yepez, E., Tarin, T., Robles-Zazueta, C., Garatuza-Payan, J., Rodriguez, J., Watts, C., Vivoni, E., 2020. Understory contribution to water vapor and co2 fluxes from a subtropical shrubland in northwestern mexico. Tecnología y ciencias del agua 11, 98–135. URL: https://revistatyca.org.mx/index.php/tyca/article/view/2313, doi:10.24850/j-tyca-2020-05-04.

Tweedie, C., 2024. AmeriFlux BASE US-Jo1 Jornada Experimental Range Bajada Site, Ver. 4-5, AmeriFlux AMP, (Dataset). doi:10.17190/AMF/1767833.

Vandegehuchte, M., Steppe, K., 2013. Sap-flux density measurement methods: Working principles and applicability. Functional Plant Biology 40, 213–223. doi:10.1071/FP12233.

Vivoni, E., Aragon, C., Malczynski, L., Tidwell, V., 2009. Semiarid watershed response in central new mexico and its sensitivity to climate variability and change. Hydrology and Earth System Sciences 13, 715–733. doi:10.5194/hess-13-715-2009.

Vivoni, E., Moreno, H., Mascaro, G., Rodriguez, J., Watts, C., Garatuza-Payan, J., Scott, R., 2008. Observed relation between evapotranspiration and soil moisture in the north american monsoon region. Geophys. Res. Lett. 35, L22403. doi:10.1029/2008GL036001.

Waring, R.H., Schlesinger, W.H., 1985. Forest Ecosystems: Concepts and Management. Academic Press Inc., Orlando, San Diego. URL: https://www.google.com/books/edition/Forest_Ecosystems/EoXwAAAAMAAJ?hl=en.

West, A.G., Hultine, K.R., Jackson, T.L., Ehleringer, J.R., 2007. Differential summer water use by pinus edulis and juniperus osteosperma reflects contrasting hydraulic characteristics. Tree Physiology 27, 1711–1720. URL: https://doi.org/10.1093/treephys/27.12.1711, doi:10.1093/treephys/27.12.1711.

Williams, D., Cable, W., Hultine, K., Hoedjes, J., Yepez, E., Simonneaux, V., Er-Raki, S., Boulet, G., de Bruin, H., Chehbouni, A., Hartogensis, O., Timouk, F., 2004. Evapotranspiration components determined by stable isotope, sap flow and eddy covariance techniques. Agricultural and Forest Meteorology 125, 241–258. doi:10.1016/j.agrformet.2004.04.008.

Wilson, K., Baldocchi, D., 2001. Comparing independent estimates of carbon dioxide exchange over five years at a deciduous forest in the southern united states. Journal of Geophysical Research 106, 34167–34178. doi:10.1029/2001JD000624.

Wilson, K.B., Baldocchi, D.D., Hanson, P.J., 2001. Leaf age affects the seasonal pattern of photosynthetic capacityand net ecosystem exchange of carbon in a deciduous forest. Plant, Cell & Environment 24, 571–583. URL: https://onlinelibrary.wiley.com/doi/abs/10.1046/j.0016-8025.2001.00706.x, doi:https://doi.org/10.1046/j.0016-8025.2001.00706.x.

Yang, H., Yang, D., Lei, Z., Sun, F., 2008. New analytical derivation of the mean annual water-energy balance equation. Water Resources Research 44, W03410. URL: https://agupubs.onlinelibrary.wiley.com/doi/abs/10.1029/2007WR006135, doi:https://doi.org/10.1029/2007WR006135.

Yepez, E.A., Huxman, T.E., Ignace, D.D., English, N.B., Weltzin, J.F., Castellanos, A.E.,

Williams, D.G., 2005. Dynamics of transpiration and evaporation following a moisture pulse in semiarid grassland: A chamber-based isotope method for partitioning flux components. Agricultural and Forest Meteorology 132, 359–376. doi:10.1016/j.agrformet.2005.09.006.

Yue, G., Zhao, H., Zhang, T.H., Zhao, X., Niu, L., Drake, S., 2008. Evaluation of water use of caragana microphylla with the stem heat-balance method in horqin sandy land, inner mongolia, china. Agricultural and Forest Meteorology 148, 1668–1678. doi:10.1016/j.agrformet.2008.05.019.

Zahn, E., Bou-Zeid, E., Good, S.P., Katul, G.G., Thomas, C.K., Ghannam, K., Smith, J.A., Chamecki, M., Dias, N.L., Fuentes, J.D., Alfieri, J.G., Kwon, H., Caylor, K.K., Gao, Z., Soderberg, K., Bambach, N.E., Hipps, L.E., Prueger, J.H., Kustas, W.P., 2022. Direct partitioning of eddy-covariance water and carbon dioxide fluxes into ground and plant components. Agricultural and Forest Meteorology 315, 108790. URL: https://www.sciencedirect.com/science/article/pii/S0168192321004767, doi:https://doi.org/10.1016/j.agrformet.2021.108790.

Zeng, X., Dickinson, R., Walker, A., Shaikh, M., Defries, R., Qi, J., 2000. Derivation and evaluation of global 1-km fractional vegetation cover data for land modeling. Journal of Applied Meteorology 39, 826–839. doi:10.1175/1520-0450(2000)039<0826:DAEOGK>2.0.CO;2.

Zhang, L., Dawes, W.R., Walker, G.R., 2001. Response of mean annual evapotranspiration to vegetation changes at catchment scale. Water Resources Research 37, 701–708. URL: https://agupubs.onlinelibrary.wiley.com/doi/abs/10.1029/2000WR900325, doi:https://doi.org/10.1029/2000WR900325.

Zhang, Y., Chiew, F.H.S., Peña-Arancibia, J., Sun, F., Li, H., Leuning, R., 2017. Global variation of transpiration and soil evaporation and the role of their major climate drivers. Journal of Geophysical Research: Atmospheres 122, 6868–6881. URL: https://agupubs.onlinelibrary.wiley.com/doi/abs/10.1002/2017JD027025, doi:https://doi.org/10.1002/2017JD027025.

Zhou, S., Yu, B., Zhang, Y., Huang, Y., Wang, G., 2016. Partitioning evapotranspiration based on the concept of underlying water use efficiency. Water Resources Research 52, 1160–1175. URL: https://agupubs.onlinelibrary.wiley.com/doi/abs/10.1002/2015WR017766, doi:https://doi.org/10.1002/2015WR017766.